\documentclass[aps,prl,twocolumn,groupedaddress,notitlepage,
floatfix,superscriptaddress]{revtex4-2}

\pdfoutput=1 
\usepackage{graphicx,graphics,epsfig,subfigure,times,bm,bbm,amssymb,amsmath,amsthm,mathrsfs,MnSymbol} \usepackage{gensymb} \usepackage{amsfonts} \usepackage{float} \usepackage[matrix,frame,arrow]{xypic} \usepackage[pdfstartview=FitH]{hyperref} 
\usepackage{times}
\usepackage{float}
\usepackage{graphics}
\usepackage[T1]{fontenc}
\usepackage{CJKutf8}
\usepackage{braket}  
\usepackage{enumerate} \usepackage[normalem]{ulem}
\usepackage[usenames,dvipsnames]{xcolor} \usepackage{multirow} \usepackage{mathtools}
\usepackage{bbm} \usepackage{titletoc} 

\hypersetup{ colorlinks=true,       
	linkcolor=red,          
	citecolor=blue,        
	filecolor=magenta,      
	urlcolor=blue,           
	runcolor=cyan }

\newcommand{\beginsupplement}{%
  \renewcommand\figurename{Supplementary Figure}
  \renewcommand\tablename{Supplementary Table}
  \setcounter{table}{0}
  \renewcommand{\thetable}{~\arabic{table}}%
  \setcounter{figure}{0}
  \renewcommand{\thefigure}{~\arabic{figure}}%
  \setcounter{section}{0}
  \setcounter{equation}{0}
  \setcounter{page}{1}
}

\begin{document}

	\title{Observation of Inelastic Meson Scattering in a Floquet System using a Digital Quantum Simulator}
	
	\author{Ziting Wang}
		\altaffiliation[]{These authors contributed equally to this work.}
	   \affiliation{Beijing Key Laboratory of Fault-Tolerant Quantum Computing, Beijing Academy of Quantum Information Sciences, Beijing 100193, China} 
	
	\author{Zi-Yong~Ge}
		\altaffiliation[]{These authors contributed equally to this work.}
		\affiliation{Quantum Information Physics Theory Research Team, Center for Quantum Computing, RIKEN, Wako-shi, Saitama 351-0198, Japan}
		
   \author{Yun-Hao Shi}
   \altaffiliation[]{These authors contributed equally to this work.}
	\affiliation{Institute of Physics, Chinese Academy of Sciences, Beijing 100190, China}
	
	\author{Zheng-An Wang}
	  \affiliation{Beijing Key Laboratory of Fault-Tolerant Quantum Computing, Beijing Academy of Quantum Information Sciences, Beijing 100193, China} 
	  
	  \author{Si-Yun Zhou}
	  	 \affiliation{Institute of Physics, Chinese Academy of Sciences, Beijing 100190, China} 
	  \affiliation{School of Physical Sciences, University of Chinese Academy of Sciences, Beijing 100190, China} 
	  
	  	  \author{Hao Li}
	  \affiliation{Beijing Key Laboratory of Fault-Tolerant Quantum Computing, Beijing Academy of Quantum Information Sciences, Beijing 100193, China} 
	  
	   \author{Kui Zhao}
	  \affiliation{Beijing Key Laboratory of Fault-Tolerant Quantum Computing, Beijing Academy of Quantum Information Sciences, Beijing 100193, China} 
	  
	   \author{Yue-Shan Xu}
	  \affiliation{Beijing Key Laboratory of Fault-Tolerant Quantum Computing, Beijing Academy of Quantum Information Sciences, Beijing 100193, China} 
	  
	   \author{Wei-Guo Ma}
	 	 \affiliation{Institute of Physics, Chinese Academy of Sciences, Beijing 100190, China} 
	 \affiliation{School of Physical Sciences, University of Chinese Academy of Sciences, Beijing 100190, China} 
	  
	    \author{Hao-Tian Liu}
	 	 \affiliation{Institute of Physics, Chinese Academy of Sciences, Beijing 100190, China} 
	 \affiliation{School of Physical Sciences, University of Chinese Academy of Sciences, Beijing 100190, China} 
	  
	    	  \author{Cai-Ping Fang}
	 	 \affiliation{Institute of Physics, Chinese Academy of Sciences, Beijing 100190, China} 
	 \affiliation{School of Physical Sciences, University of Chinese Academy of Sciences, Beijing 100190, China} 
	  
	    	  \author{Jia-Cheng Song}
		 \affiliation{Institute of Physics, Chinese Academy of Sciences, Beijing 100190, China} 
	\affiliation{School of Physical Sciences, University of Chinese Academy of Sciences, Beijing 100190, China} 
	  
	    	  \author{Tian-Ming Li}
	  	 \affiliation{Institute of Physics, Chinese Academy of Sciences, Beijing 100190, China} 
	  \affiliation{School of Physical Sciences, University of Chinese Academy of Sciences, Beijing 100190, China} 
	  
	    	  \author{Jiachi Zhang }
		 \affiliation{Institute of Physics, Chinese Academy of Sciences, Beijing 100190, China} 
	\affiliation{School of Physical Sciences, University of Chinese Academy of Sciences, Beijing 100190, China} 
	  
	    	  \author{Yu Liu}
	  \affiliation{Institute of Physics, Chinese Academy of Sciences, Beijing 100190, China} 
	  
	    	  \author{Cheng-Lin Deng}
	  \affiliation{Beijing Key Laboratory of Fault-Tolerant Quantum Computing, Beijing Academy of Quantum Information Sciences, Beijing 100193, China}

    \author{Guangming Xue}
	   \affiliation{Beijing Key Laboratory of Fault-Tolerant Quantum Computing, Beijing Academy of Quantum Information Sciences, Beijing 100193, China}

    \author{Haifeng Yu}
	   \affiliation{Beijing Key Laboratory of Fault-Tolerant Quantum Computing, Beijing Academy of Quantum Information Sciences, Beijing 100193, China} 

	\author{Kai~Xu} 
	 \affiliation{Institute of Physics, Chinese Academy of Sciences, Beijing 100190, China} 
      \affiliation{School of Physical Sciences, University of Chinese Academy of Sciences, Beijing 100190, China} 
     \affiliation{Beijing Key Laboratory of Fault-Tolerant Quantum Computing, Beijing Academy of Quantum Information Sciences, Beijing 100193, China} 
      \affiliation{Songshan Lake Materials Laboratory, Dongguan 523808, China}
      \affiliation{Hefei National Laboratory, Hefei 230088, China}

   \author{Kaixuan Huang}
	   \email{huangkx@baqis.ac.cn}
      	   \affiliation{Beijing Key Laboratory of Fault-Tolerant Quantum Computing, Beijing Academy of Quantum Information Sciences, Beijing 100193, China} 
      	   
	\author{Franco Nori}
		\email{fnori@riken.jp}
	\affiliation{Quantum Information Physics Theory Research Team, Center for Quantum Computing, RIKEN, Wako-shi, Saitama 351-0198, Japan}
	\affiliation{Department of Physics, University of Michigan, Ann Arbor, Michigan 48109-1040, USA}

	\author{Heng~Fan} 
	\email{hfan@iphy.ac.cn}
	 \affiliation{Institute of Physics, Chinese Academy of Sciences, Beijing 100190, China} 
    \affiliation{School of Physical Sciences, University of Chinese Academy of Sciences, Beijing 100190, China} 
    \affiliation{Beijing Key Laboratory of Fault-Tolerant Quantum Computing, Beijing Academy of Quantum Information Sciences, Beijing 100193, China} 
   \affiliation{Songshan Lake Materials Laboratory, Dongguan 523808, China}
	\affiliation{Hefei National Laboratory, Hefei 230088, China}

\begin{abstract}
Lattice gauge theories provide a non-perturbative framework for understanding confinement and hadronic physics, 
but their real-time dynamics remain challenging for classical computations. 
However, quantum simulators offer a promising alternative for exploring such dynamics beyond classical capabilities. 
Here, we experimentally investigate meson scattering using a superconducting quantum processor. 
Employing a digital protocol, we realize a Floquet spin chain equivalent to a one-dimensional Floquet $\mathbb{Z}_2$ lattice gauge theory.
We observe Bloch oscillations of single kinks and strong binding between adjacent kinks, signaling confinement and the formation of stable mesons in this Floquet system. 
Using full-system joint readout, we resolve meson populations by string length, enabling identification of meson scattering channels. 
Our results reveal the fragmentation of a long-string meson into multiple short-string mesons, which is also an experimental signature of string breaking.
 Moreover, we directly observe inelastic meson scattering, where two short-string mesons can merge into a longer one.
Our results pave the way for studying interacting gauge particles and composite excitations on digital quantum simulators.
\end{abstract}
	
	\maketitle
	
   \textit{Introduction.}---%
   Quarks cannot exist in isolation at low energies due to confinement; instead, they tend to form hadrons such as mesons.
   To describe this fundamental phenomenon, lattice gauge theories (LGTs) were introduced as a non-perturbative framework for gauge fields~\cite{PhysRevD.10.2445,RevModPhys.51.659}.
   While LGTs have provided profound theoretical insights, 
   simulating their non-equilibrium dynamics~\cite{ISI:000349934700015,RevModPhys.83.863}  remains a major challenge for classical computational methods~\cite{PhysRevLett.94.170201}.
   Recent advances in quantum simulation~\cite{Buluta108,RevModPhys.86.153,cheng2023noisy} offer a new route to studying LGTs in synthetic quantum many-body systems~\cite{jordan2012quantum,doi:10.1080/00107514.2016.1151199,banuls2020simulating,davoudi2022quantum,bauer2023quantum,halimeh2024universal}.
   
   In particular, various LGTs have been experimentally implemented across different platforms, 
   including $U(1)$ quantum link models~\cite{ISI:000591335100019,doi:10.1126/science.abl6277,zhang2024observation,gonzalez2025observation} and $\mathbb{Z}_2$ gauge theories~\cite{ISI:000494944200023,ISI:000494944200024,PhysRevResearch.4.L022060,mildenberger2025confinement,de2024observation,cochran2025visualizing}.
   Moreover, hallmark phenomena, such as confinement~\cite{PhysRevResearch.4.L022060,mildenberger2025confinement,ISI:000395814000014}, string breaking~\cite{PhysRevLett.111.201601,de2024observation,cochran2025visualizing}, and meson-like excitations~\cite{PhysRevLett.115.100501,PhysRevX.13.031017}
   have been  observed through real-time dynamics in these systems.
   These developments highlight the potential of quantum simulations to explore LGTs beyond the capabilities of classical approaches, 
   particularly in the non-equilibrium regime.

   Beyond static properties and confinement, hadron scattering~\cite{ellis2003qcd,ACHENBACH2024122874} is another essential aspect of gauge dynamics and plays a central role in understanding the Standard Model. However, simulating such processes is also a hard problem for classical computation~\cite{PhysRevLett.94.170201}. 
   Recent theoretical and experimental studies have shown that certain spin models can serve as effective realizations of LGTs~\cite{ISI:000395814000014,PhysRevLett.124.120503}, capturing key features, such as confinement, string breaking, and meson-like excitations. 
   These simplified models provide a highly tractable setting for studying real-time gauge dynamics and can be naturally implemented in noisy intermediate-scale quantum (NISQ) systems~\cite{PhysRevLett.122.150601}.
   A notable example is the one-dimensional Ising model with both transverse and longitudinal fields~\cite{ISI:000395814000014}, where kinks experience a confining linear potential, forming stable mesonic bound states connected by strings of anti-aligned spins. 
   While previous quantum simulation experiments have successfully probed confinement and string breaking in such systems, the real-time dynamics of meson–meson  scattering, specifically the inelastic scattering, still remains challenging  in experiments~\cite{surace2021scattering,PRXQuantum.3.040309,PhysRevResearch.4.L032001,PRXQuantum.5.020311,kreshchuk2023simulating,Chai2025fermionicwavepacket,PhysRevD.109.114510,Davoudi2024scatteringwave,PRXQuantum.5.040310,Bennewitz2025simulatingmeson,schuhmacher2025observation,joshi2025probing}.
   
   \begin{figure*}[t] \includegraphics[width=1\textwidth]{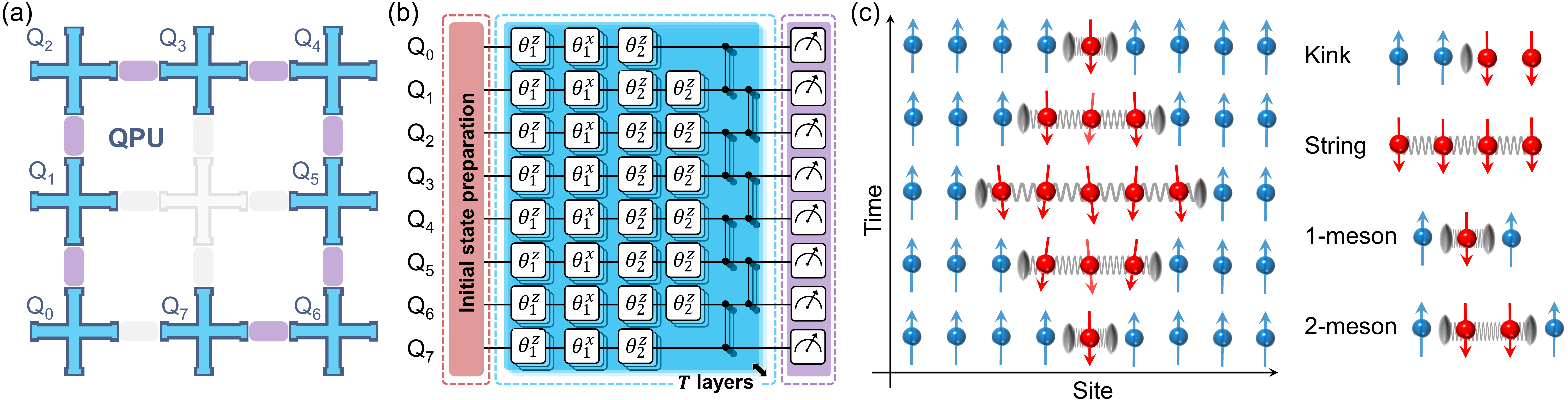}
	\caption{Set-up. (a) Diagram of the superconducting quantum processor. 
		The device consists of 9 tunable transmon qubits (blue crosses) arranged in a $3 \times 3$ square lattice, 
		with each pair of nearest-neighbor qubits connected by a tunable coupler (purple bonds).
		The light-colored qubit and couplers are used in this experiment, effectively reducing the system to a spin chain with open boundary conditions.
		(b) Quantum circuit for the digital simulation. The dynamics consists of $T$ driving cycles, each comprising three layers: the first layer applies single-qubit rotations around the $x$-axis to implement a transverse field; the second layer applies $z$-axis rotations to implement a longitudinal field; the final layer combines additional $z$-rotations and controlled-phase gates to realize the Ising interaction. A detailed gate decomposition is provided in the Supplemental Material (SM)~\cite{SM}.
		(c) Schematic illustration of kink confinement and mesons. The system is initialized with two adjacent kinks connected by a string. 
		During time evolution, the kinks remain close due to confinement, forming a bound state, i.e., a meson. 
		A meson with string length $\ell$ is labeled as an $\ell$-meson.
	}
	\label{fig_1}
\end{figure*} 

In this Letter, we experimentally investigate meson scattering using a superconducting quantum processor.
We digitally implement a Floquet spin chain, which is equivalent to a 1D Floquet $\mathbb{Z}_2$ LGT.
We first observe Bloch oscillations of a single-kink initial state and find that two adjacent kinks remain tightly bound, 
indicating the existence of confinement and stable meson excitations.
We then explore meson scattering by tracking the population of each meson with different string lengths.
Our results show the fragmentation of a long-string meson into multiple short-string mesons, which is also an experimental signature of string breaking.
Furthermore, we directly observe\textit{ inelastic meson} scattering  for a specific longitudinal field, where two short-string mesons merge into a longer one.
Our results pave the way for simulating hadron scattering on digital quantum platforms, and offer deep insights into the nonequilibrium dynamics of LGTs.

\textit{Set-up.}---%
The experiment is performed on a 2D superconducting quantum processor consisting of nine transmon qubits~\cite{koch2007charge,annurev-conmatphys-031119-050605,siddiqi2021engineering} arranged in a $3 \times 3$ square lattice,
fabricated using flip-chip technology~\cite{Google_EC_2025} with a tantalum base layer~\cite{fab,fab2},
see Fig.~\ref{fig_1}(a). 
Each nearest-neighbor pair is coupled via a tunable coupler that enables high-fidelity two-qubit gates. 
The median energy relaxation time $T_1$ is approximately $30~\mu\mathrm{s}$. 
Detailed device parameters are provided in the Supplemental Material (SM)~\cite{SM} and Refs.~\cite{wang_quantum_2024,wang_demonstration_2024}. 
In this work, we use 8 qubits along the edge of the lattice to form a 1D spin chain, as illustrated in Fig.~\ref{fig_1}(a).

\begin{figure*}[t] \includegraphics[width=1\textwidth]{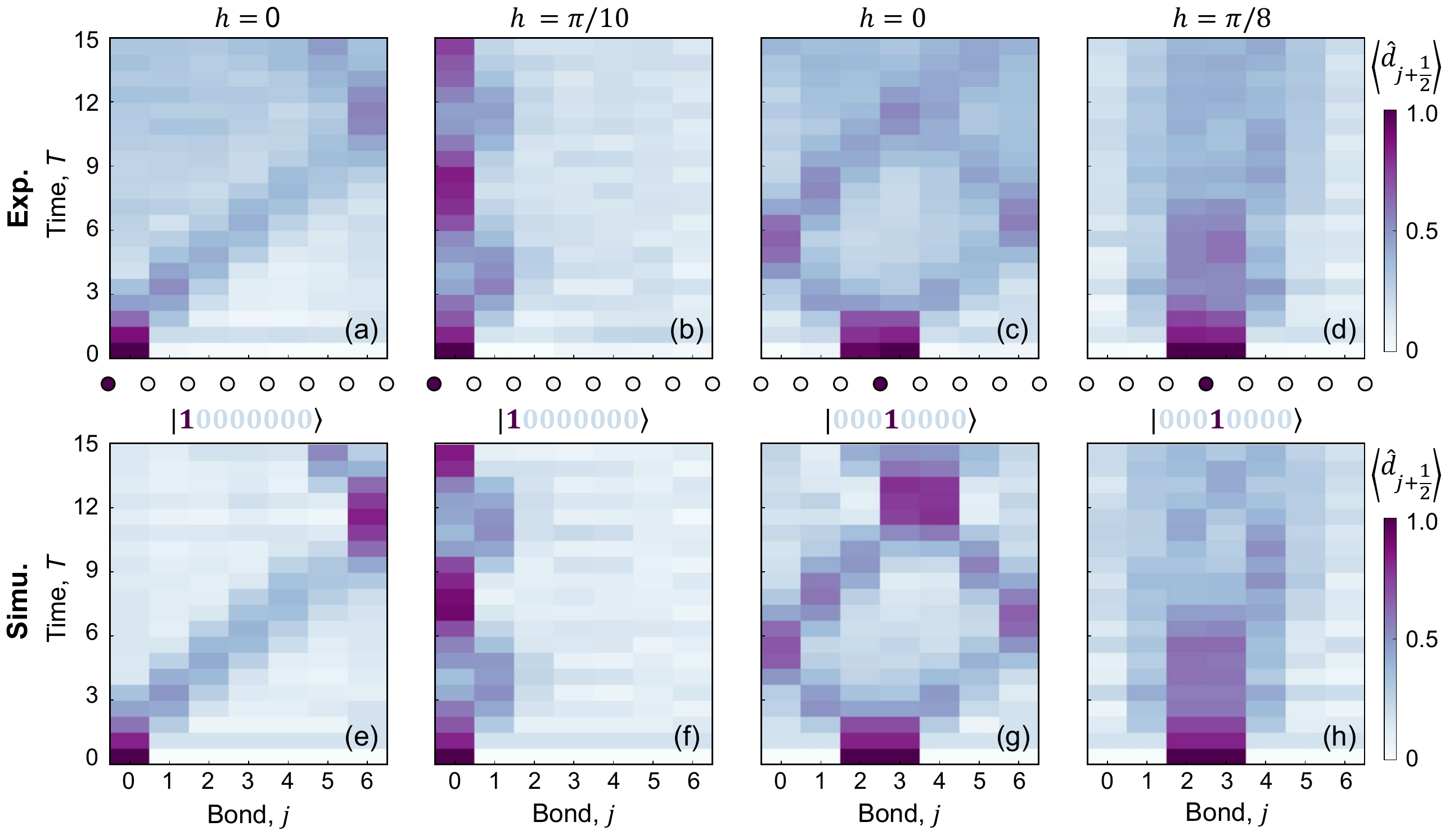}
   \caption{Confinement and meson excitations. 
   	The dynamics of kink distributions for the single-kink initial state $\ket{\psi_0} = \ket{10000000}$ with (a) $h=0$ and (b) $h=\pi/10$.
   	The dynamics of kink distributions for the single-meson initial state $\ket{\psi_1} = \ket{00010000}$ with (c) $h=0$ and (d) $h=\pi/8$.
   	(e--h) Numerical results of (a--d) using the ideal quantum circuits.}
   \label{fig_2}
\end{figure*} 

   We consider a Floquet system where each driving period is governed by the unitary operator
   \begin{align} \label{U}
   \hat{U} = \exp(-i h \hat{Z})\exp(-i \mu \hat{X}) \exp(i J \hat{H}_{zz}),
   \end{align}
   with $\hat{Z} = \sum_j \hat{\sigma}_j^z$, $\hat{X} = \sum_j \hat{\sigma}_j^x$, and $\hat{H}_{zz} = -\sum_j \hat{\sigma}_j^z \hat{\sigma}_{j+1}^z$, where $\hat{\sigma}_j^\alpha$ ($\alpha = x, y, z$) denote the Pauli matrices. The Floquet evolution $\hat{U}$ is implemented digitally: single-qubit rotations $e^{-i \mu \hat{X}}$ and $e^{-i h \hat{Z}}$ are realized with a median fidelity $99.95\% $, while the entangling evolution $e^{i J \hat{H}_{zz}}$ is realized via controlled-phase gates with median fidelity $99.33\%$~\cite{SM}. The full quantum circuit is shown in Fig.~\ref{fig_1}(b), where the evolution time is characterized by total cycles $T$. 
   In this experiment, the maximum $T$ is up to 15, where the finial state still retains high fidelity.

   In the perturbative regime $h, \mu, J \ll 1$, the Floquet dynamics approximate those of a time-independent Ising model with transverse and longitudinal fields. 
   This model is equivalent to a $\mathbb{Z}_2$ LGT coupled to matter fields and serves as a paradigmatic setting for exploring confinement and meson excitations~\cite{ISI:000395814000014}.
   For $h = 0$ and $J > \mu$, the system resides in a ferromagnetic phase, where the elementary excitations are kinks (domain walls). 
   When $h \ne 0$, kinks become confined, and tend to form bound states connected by strings of anti-aligned spins, i.e., mesons [see Fig.~\ref{fig_1}(c)].

   In this experiment, we explore a nonperturbative regime with $J = \pi/4$ and $\mu = \pi/10$, 
   where the dynamics deviate  from those generated by a time-independent Hamiltonian.
   However, we can still use a Floquet $\mathbb{Z}_2$ LGT coupled to a matter field to describe the dynamics as~\cite{10.21468/SciPostPhys.10.6.148,PhysRevLett.124.120503,SM}
      \begin{align} \label{Ulgt}\nonumber
   	\hat{U} = &\exp({-i h\sum_j \hat{\tau}_{j+1/2}^x})\exp( {-i\mu\sum_{j}\hat s_j^x \hat \tau^z_{j+\frac{1}{2}} \hat s_{j+1}^x})\\
   &	\exp( iJ\sum_{j}\hat s_j^z),
   \end{align}
   where  $\hat{s}_j^\alpha$ and $\hat{\tau}_j^\alpha$ ($\alpha = x, y, z$) are both the Pauli matrices, representing the matter and gauge fields, respectively.
   Here, the  $\mathbb{Z}_2$ gauge generator is $\hat G_{j} = \hat \tau^x_{j-\frac{1}{2}} \hat s_j^z\hat \tau^x_{j+\frac{1}{2}} $, i.e., $ [\hat G_j, \hat  U ]=0$.
   In addition, for $h = 0$ and $J > \mu$, the system remains in a Floquet ferromagnetic phase supporting kink excitations~\cite{PhysRevLett.116.250401,PhysRevB.93.245146,annurev:/content/journals/10.1146/annurev-conmatphys-031218-013721}. 
   Thus, when a finite longitudinal field is introduced, confinement and stable meson excitations are expected to emerge in this Floquet system. 
   In the following, we will experimentally demonstrate this conjecture and investigate the meson scattering. 
   Here, we call a meson as a $\ell$-meson, when its string length is $\ell$, see Fig.~\ref{fig_1}(c).
   We also note that these mesons are bare mesons, corresponding to the exact eigenstates of the $\mu = 0$ (non-interacting) limit.

  \textit{Confinement and mesons.}---%
  We first demonstrate confinement dynamics and meson excitations in this Floquet system. 
  As a starting point, we prepare a single-kink initial state, $\ket{\psi_0} = \ket{10000000}$, and monitor the kink density, defined as
    \begin{align} \label{dw}
		\hat{d}_{j+\frac{1}{2}}= (1-\hat \sigma_j^z \hat \sigma_{j+1}^z)/2.
	 \end{align} 
  For $h = 0$, the kink propagates ballistically and forms a light-cone-like profile [see Fig.~\ref{fig_2}(a)], indicating that kinks are free excitations in this case. 
  In contrast, for a finite longitudinal field $h = \pi/10$, the kink remains localized and exhibits an oscillation near its initial position [see Fig.~\ref{fig_2}(b)]. 
  This behavior resembles Bloch oscillations~\cite{RevModPhys.34.645,PhysRevB.46.7252,PhysRevLett.76.4508,guo2021observation}, arising from the linear potential introduced by the longitudinal field.
  Therefore, in this single-kink system, the emergence of Bloch oscillations indicates the existence of effective linear potential for kinks~\cite{ISI:000395814000014},
  which is strong evidence of confinement.

We next consider an initial state with two adjacent kinks, i.e., a 1-meson, given by $\ket{\psi_1} = \ket{00010000}$. 
When $h = 0$, the two kinks separate and propagate freely over long distances [see Fig.~\ref{fig_2}(c)],
indicating the absence of confinement and the instability of mesons. 
However, for $h = \pi/8$, the two kinks remain tightly bound  as a composite excitation throughout the evolution [see Fig.~\ref{fig_2}(d)]. 
This observation confirms the confinement and existence of stable mesons  in this Floquet system.
We also perform numerical simulations for the ideal quantum circuits, with results which are consistent with our experimental data [Figs.~\ref{fig_2}(e--h)].

\begin{figure}[t] \includegraphics[width=0.48\textwidth]{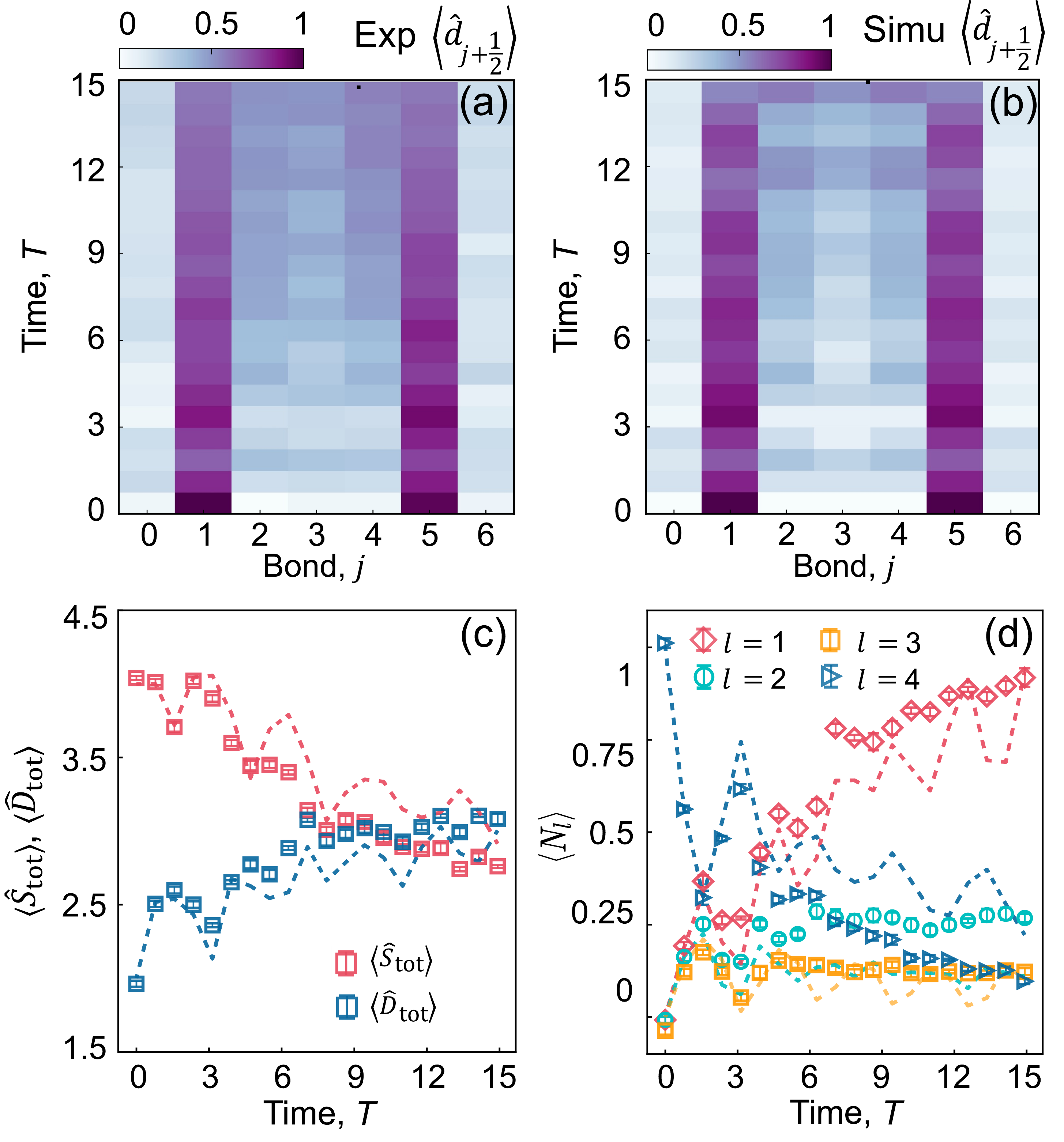}
	\caption{Fragmentation of a 4-meson for the initial state $\ket{\psi_2} = \ket{00111100}$ and $h=\pi/4$.
  (a) Experimental and (b) numerical results of kink dynamics. (c) Total  spin flips $\hat{S}_{\mathrm{tot}}$ and the total kink number $\hat{D}_{\mathrm{tot}}$ versus evolution time.
  (d) The dynamics of the  meson populations with different string length $\braket{\hat{N}_\ell}$.
  The dashed curves represent the corresponding numerical results.
  The error bars are obtained as the variance of six groups of data.
	 }
	\label{fig_3}
\end{figure}

    \textit{Meson scattering.}---%
    Having established the existence of stable meson excitations in the Floquet system for $h \ne 0$, we now investigate meson–meson scattering. 
    We first consider the dynamics of a long-string meson, where we choose the initial state $\ket{\psi_2} = \ket{00111100}$,
     i.e., there exists a 4-meson. 
    The dynamics of the kink density $\hat{d}_{j+1/2}$ for $h = \pi/4$ is shown in Figs.~\ref{fig_3}(a,b). 
    We observe that kinks tend to distribute within the region between the initial kinks (i.e., between qubits $Q_2$ and $Q_5$) after a long-time evolution, 
    indicating the emergence of short-string mesons.
    To analyze this process quantitatively, we also measure the total spin flips $\hat{S}_{\mathrm{tot}} = \sum_j \hat{\sigma}_j^+ \hat{\sigma}_j^-$ and the total kink number $\hat{D}_{\mathrm{tot}} = \sum_{j} \hat{d}_{j+1/2}$. 
    As shown in Fig.~\ref{fig_3}(c), $\langle \hat{S}_{\mathrm{tot}} \rangle$ decreases, while $\langle \hat{D}_{\mathrm{tot}} \rangle$ increases, with the time evolution, 
    indicating that the initial 4-meson decays into multiple shorter-string mesons,
    which can also be understood as string breaking.

    To further identify the scattering channel, we track the population of $\ell$-mesons, defined as
    \begin{align} \label{ml}
   	    	\hat{M}_{j,\ell} := \hat P^0_{j-1}\bigg(\prod_{k=j}^{k=j+\ell}\hat P^1_k\bigg)\hat P^0_{j+\ell+1},
   	    \end{align}
    where $\hat{P}^0 = \ket{0}\bra{0}$ and $\hat{P}^1 = \ket{1}\bra{1}$ are projection operators. 
    The total number of $\ell$-mesons is then given by  $\hat{N}_\ell = \sum_j \hat{M}_{j,\ell}$. 
    In the experiment, the expectation value of $\hat{M}_{j,\ell}$ can be measured by joint readout of all qubits.
    As shown in Fig.~\ref{fig_3}(d), the population of the 4-meson indeed decreases over time, accompanied by the emergence of  short-string mesons,
    especially 1-mesons. 
    This result directly demonstrates a scattering process: a 4-meson fragments into multiple 1- mesons.

\begin{figure}[t] \includegraphics[width=0.48 \textwidth]{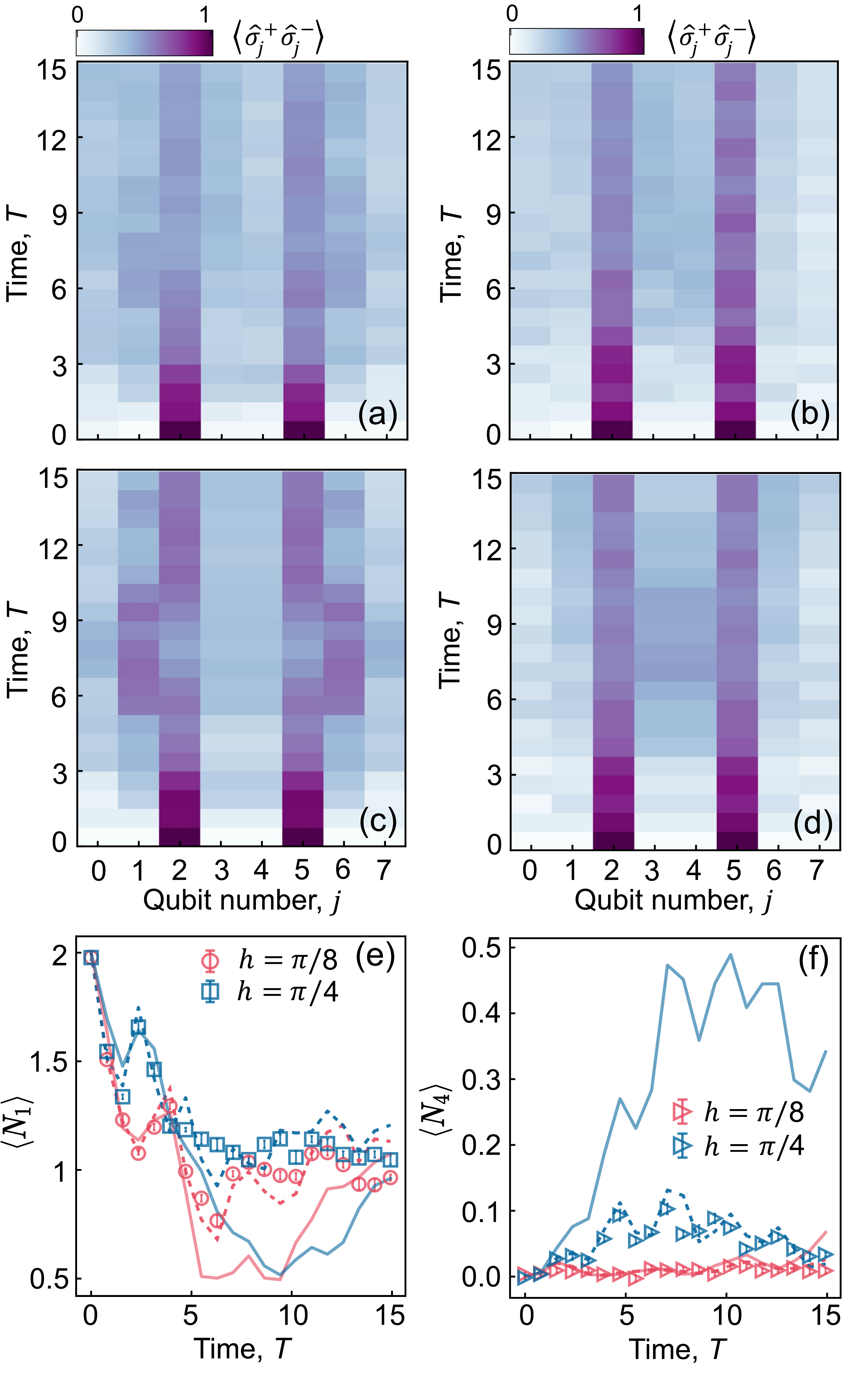}
	\caption{ Scattering between two 1-mesons for the initial state $\ket{\psi_3} = \ket{00100100}$.
	Experimental results of dynamics of spin flips  $\braket{\hat\sigma_j^+\hat\sigma^-_j}$ with (a) $h=\pi/8$ and (b) $h=\pi/4$. 
	(c,d) Corresponding numerical simulations of the ideal quantum circuits for (a) and (b), respectively.
	(e,f) Time evolution of the populations of 1-mesons and 4-mesons, respectively.
	Error bars represent the standard deviation over six independent measurement groups.
	Dashed curves show numerical results from the ideal quantum circuits, 
	while solid curves correspond to numerical simulations using the time-dependent Hamiltonian in Eq.~(\ref{H}) with the same parameters.
        }
	\label{fig_4}
\end{figure}

    We further investigate the scattering between two 1-mesons using the initial state $\ket{\psi_3} = \ket{00100100}$.
    Although these two mesons are  localized, their proximity allows for interaction and possible scattering.
    We focus on two values of the longitudinal field, $h=\pi/8$ and $h=\pi/4$, 
    and plot the spin-flip dynamics $\braket{\hat \sigma^+_j\hat \sigma^-_j}$ in Figs.~\ref{fig_4}(a–d).
    For $h=\pi/4$, spin flips are more pronounced between qubits $Q_2$ and $Q_5$, suggesting the formation of more longer-string mesons.
    To further characterize the meson distribution, we track the population of $\ell$-mesons by measuring  $\hat{N}_\ell$.
    For both values of $h$, the initial 1-mesons decay [Fig.\ref{fig_4}(e)].
    However, 4-mesons emerge only for $h=\pi/4$ [Fig.\ref{fig_4}(f)].
    This provides strong evidence for inelastic meson scattering~\cite{surace2021scattering,PRXQuantum.3.040309,PhysRevResearch.4.L032001,Bennewitz2025simulatingmeson}, where two short-string mesons merge into a longer one.
    
   We also numerically simulate meson dynamics under a time-independent Hamiltonian
      \begin{align} \label{H}
   	\hat H = -J\sum_{j}\hat\sigma_j^z\hat\sigma_{j+1}^z + \mu\sum_{j}\hat\sigma_j^x + h\sum_{j}\hat\sigma_j^z,
   \end{align}
   with $J=\pi/4$, $\mu=\pi/10$, and the same initial state $\ket{\psi_3} = \ket{00100100}$.
   Similar to the Floquet case, inelastic scattering occurs for $h=\pi/4$, as shown in Figs.~\ref{fig_4}(e,f).
    In addition, the 4-meson population is  higher in this time-independent Hamiltonian, 
    indicating that the Floquet drive can suppress inelastic meson scattering.

    \textit{Summary}---%
    In conclusion, we experimentally investigate meson scattering in a superconducting quantum processor by digitally simulating a Floquet spin chain, 
    which can be described by a 1D Floquet $\mathbb{Z}_2$ LGT. 
    We observe confinement and stable meson excitations through real-time kink dynamics.
    Moreover, we demonstrate  the fragmentation of long-string mesons, 
    and also reveal inelastic meson scattering, where two short-string mesons merge into a longer one for a specific longitudinal field. 
    Our work provide strong experimental evidence of meson–meson inelastic scattering in a spin model, 
    and paves the way for simulating interacting gauge particles and complex non-equilibrium phenomena of LGTs in digital quantum simulators. 
    Here we mainly consider localized bare mesons, so it will be interesting to study the scattering of dressed mesons with finite momentum~\cite{surace2021scattering,PRXQuantum.3.040309,PhysRevResearch.4.L032001,PRXQuantum.5.020311,kreshchuk2023simulating,Chai2025fermionicwavepacket,PhysRevD.109.114510,Davoudi2024scatteringwave,PRXQuantum.5.040310,Bennewitz2025simulatingmeson,schuhmacher2025observation,joshi2025probing}.
    In addition, exploring meson-related (non-)thermal dynamics presents another intriguing direction for future research~\cite{PhysRevX.10.021041,PhysRevLett.124.207602,PhysRevLett.118.266601,PhysRevLett.118.266601,PhysRevLett.132.230403}.

	\textit{Acknowledgments.}---%
	K.H. is supported by National Natural Science Foundation of China (Grants No.12404578) and
Beijing National Laboratory for Condensed Matter Physics (2024BNLCMPKF022). H.F. and K.X. are supported in part by the National Natural Science Foundation of China (Grants Nos. 92265207, T2121001,
92365301, T2322030), the Innovation Program for Quantum Science and Technology (Grant No. 2021ZD0301800), the Beijing Nova Program (No. 20220484121, 20240484652). Y.-H.S. is supported by the China Postdoctoral Science Foundation (Grant No. GZB20240815). F.N. is supported in part by: the Japan Science and Technology Agency (JST) [via the CREST Quantum Frontiers program Grant No. JPMJCR24I2, the Quantum Leap Flagship Program (Q-LEAP), and the Moonshot R\&D Grant Number JPMJMS2061].

\bibliographystyle{apsrev-new}
\bibliography{ref_arxiv} 	

\begin{thebibliography}{66}%
\makeatletter
\providecommand \@ifxundefined [1]{%
 \@ifx{#1\undefined}
}%
\providecommand \@ifnum [1]{%
 \ifnum #1\expandafter \@firstoftwo
 \else \expandafter \@secondoftwo
 \fi
}%
\providecommand \@ifx [1]{%
 \ifx #1\expandafter \@firstoftwo
 \else \expandafter \@secondoftwo
 \fi
}%
\providecommand \natexlab [1]{#1}%
\providecommand \enquote  [1]{``#1''}%
\providecommand \bibnamefont  [1]{#1}%
\providecommand \bibfnamefont [1]{#1}%
\providecommand \citenamefont [1]{#1}%
\providecommand \href@noop [0]{\@secondoftwo}%
\providecommand \href [0]{\begingroup \@sanitize@url \@href}%
\providecommand \@href[1]{\@@startlink{#1}\@@href}%
\providecommand \@@href[1]{\endgroup#1\@@endlink}%
\providecommand \@sanitize@url [0]{\catcode `\\12\catcode `\$12\catcode
  `\&12\catcode `\#12\catcode `\^12\catcode `\_12\catcode `\%12\relax}%
\providecommand \@@startlink[1]{}%
\providecommand \@@endlink[0]{}%
\providecommand \url  [0]{\begingroup\@sanitize@url \@url }%
\providecommand \@url [1]{\endgroup\@href {#1}{\urlprefix }}%
\providecommand \urlprefix  [0]{URL }%
\providecommand \Eprint [0]{\href }%
\providecommand \doibase [0]{https://doi.org/}%
\providecommand \selectlanguage [0]{\@gobble}%
\providecommand \bibinfo  [0]{\@secondoftwo}%
\providecommand \bibfield  [0]{\@secondoftwo}%
\providecommand \translation [1]{[#1]}%
\providecommand \BibitemOpen [0]{}%
\providecommand \bibitemStop [0]{}%
\providecommand \bibitemNoStop [0]{.\EOS\space}%
\providecommand \EOS [0]{\spacefactor3000\relax}%
\providecommand \BibitemShut  [1]{\csname bibitem#1\endcsname}%
\let\auto@bib@innerbib\@empty
\bibitem [{\citenamefont {Wilson}(1974)}]{PhysRevD.10.2445}%
  \BibitemOpen
  \bibfield  {author} {\bibinfo {author} {\bibfnamefont {K.~G.}\ \bibnamefont
  {Wilson}},\ }\bibinfo {title} {Confinement of quarks},\ \href
  {https://doi.org/10.1103/PhysRevD.10.2445} {\bibfield  {journal} {\bibinfo
  {journal} {Phys. Rev. D}\ }\textbf {\bibinfo {volume} {10}},\ \bibinfo
  {pages} {2445} (\bibinfo {year} {1974})}\BibitemShut {NoStop}%
\bibitem [{\citenamefont {Kogut}(1979)}]{RevModPhys.51.659}%
  \BibitemOpen
  \bibfield  {author} {\bibinfo {author} {\bibfnamefont {J.~B.}\ \bibnamefont
  {Kogut}},\ }\bibinfo {title} {An introduction to lattice gauge theory and
  spin systems},\ \href {https://doi.org/10.1103/RevModPhys.51.659} {\bibfield
  {journal} {\bibinfo  {journal} {Rev. Mod. Phys.}\ }\textbf {\bibinfo {volume}
  {51}},\ \bibinfo {pages} {659} (\bibinfo {year} {1979})}\BibitemShut
  {NoStop}%
\bibitem [{\citenamefont {Eisert}\ \emph {et~al.}(2015)\citenamefont {Eisert},
  \citenamefont {Friesdorf},\ and\ \citenamefont
  {Gogolin}}]{ISI:000349934700015}%
  \BibitemOpen
  \bibfield  {author} {\bibinfo {author} {\bibfnamefont {J.}~\bibnamefont
  {Eisert}}, \bibinfo {author} {\bibfnamefont {M.}~\bibnamefont {Friesdorf}},\
  and\ \bibinfo {author} {\bibfnamefont {C.}~\bibnamefont {Gogolin}},\
  }\bibinfo {title} {Quantum many-body systems out of equilibrium},\ \href
  {https://doi.org/10.1038/NPHYS3215} {\bibfield  {journal} {\bibinfo
  {journal} {Nat. Phys.}\ }\textbf {\bibinfo {volume} {11}},\ \bibinfo {pages}
  {124} (\bibinfo {year} {2015})}\BibitemShut {NoStop}%
\bibitem [{\citenamefont {Polkovnikov}\ \emph {et~al.}(2011)\citenamefont
  {Polkovnikov}, \citenamefont {Sengupta}, \citenamefont {Silva},\ and\
  \citenamefont {Vengalattore}}]{RevModPhys.83.863}%
  \BibitemOpen
  \bibfield  {author} {\bibinfo {author} {\bibfnamefont {A.}~\bibnamefont
  {Polkovnikov}}, \bibinfo {author} {\bibfnamefont {K.}~\bibnamefont
  {Sengupta}}, \bibinfo {author} {\bibfnamefont {A.}~\bibnamefont {Silva}},\
  and\ \bibinfo {author} {\bibfnamefont {M.}~\bibnamefont {Vengalattore}},\
  }\bibinfo {title} {Colloquium: Nonequilibrium dynamics of closed interacting
  quantum systems},\ \href {https://doi.org/10.1103/RevModPhys.83.863}
  {\bibfield  {journal} {\bibinfo  {journal} {Rev. Mod. Phys.}\ }\textbf
  {\bibinfo {volume} {83}},\ \bibinfo {pages} {863} (\bibinfo {year}
  {2011})}\BibitemShut {NoStop}%
\bibitem [{\citenamefont {Troyer}\ and\ \citenamefont
  {Wiese}(2005)}]{PhysRevLett.94.170201}%
  \BibitemOpen
  \bibfield  {author} {\bibinfo {author} {\bibfnamefont {M.}~\bibnamefont
  {Troyer}}\ and\ \bibinfo {author} {\bibfnamefont {U.-J.}\ \bibnamefont
  {Wiese}},\ }\bibinfo {title} {Computational Complexity and Fundamental
  Limitations to Fermionic Quantum Monte Carlo Simulations},\ \href
  {https://doi.org/10.1103/PhysRevLett.94.170201} {\bibfield  {journal}
  {\bibinfo  {journal} {Phys. Rev. Lett.}\ }\textbf {\bibinfo {volume} {94}},\
  \bibinfo {pages} {170201} (\bibinfo {year} {2005})}\BibitemShut {NoStop}%
\bibitem [{\citenamefont {Buluta}\ and\ \citenamefont
  {Nori}(2009)}]{Buluta108}%
  \BibitemOpen
  \bibfield  {author} {\bibinfo {author} {\bibfnamefont {I.}~\bibnamefont
  {Buluta}}\ and\ \bibinfo {author} {\bibfnamefont {F.}~\bibnamefont {Nori}},\
  }\bibinfo {title} {Quantum simulators},\ \href
  {https://doi.org/10.1126/science.1177838} {\bibfield  {journal} {\bibinfo
  {journal} {Science}\ }\textbf {\bibinfo {volume} {326}},\ \bibinfo {pages}
  {108} (\bibinfo {year} {2009})}\BibitemShut {NoStop}%
\bibitem [{\citenamefont {Georgescu}\ \emph {et~al.}(2014)\citenamefont
  {Georgescu}, \citenamefont {Ashhab},\ and\ \citenamefont
  {Nori}}]{RevModPhys.86.153}%
  \BibitemOpen
  \bibfield  {author} {\bibinfo {author} {\bibfnamefont {I.~M.}\ \bibnamefont
  {Georgescu}}, \bibinfo {author} {\bibfnamefont {S.}~\bibnamefont {Ashhab}},\
  and\ \bibinfo {author} {\bibfnamefont {F.}~\bibnamefont {Nori}},\ }\bibinfo
  {title} {Quantum simulation},\ \href
  {https://doi.org/10.1103/RevModPhys.86.153} {\bibfield  {journal} {\bibinfo
  {journal} {Rev. Mod. Phys.}\ }\textbf {\bibinfo {volume} {86}},\ \bibinfo
  {pages} {153} (\bibinfo {year} {2014})}\BibitemShut {NoStop}%
\bibitem [{\citenamefont {Cheng}\ \emph {et~al.}(2023)\citenamefont {Cheng},
  \citenamefont {Deng}, \citenamefont {Gu}, \citenamefont {He}, \citenamefont
  {Hu}, \citenamefont {Huang}, \citenamefont {Li}, \citenamefont {Lin},
  \citenamefont {Lu}, \citenamefont {Lu} \emph {et~al.}}]{cheng2023noisy}%
  \BibitemOpen
  \bibfield  {author} {\bibinfo {author} {\bibfnamefont {B.}~\bibnamefont
  {Cheng}}, \bibinfo {author} {\bibfnamefont {X.-H.}\ \bibnamefont {Deng}},
  \bibinfo {author} {\bibfnamefont {X.}~\bibnamefont {Gu}}, \bibinfo {author}
  {\bibfnamefont {Y.}~\bibnamefont {He}}, \bibinfo {author} {\bibfnamefont
  {G.}~\bibnamefont {Hu}}, \bibinfo {author} {\bibfnamefont {P.}~\bibnamefont
  {Huang}}, \bibinfo {author} {\bibfnamefont {J.}~\bibnamefont {Li}}, \bibinfo
  {author} {\bibfnamefont {B.-C.}\ \bibnamefont {Lin}}, \bibinfo {author}
  {\bibfnamefont {D.}~\bibnamefont {Lu}}, \bibinfo {author} {\bibfnamefont
  {Y.}~\bibnamefont {Lu}}, \emph {et~al.},\ }\bibinfo {title} {Noisy
  intermediate-scale quantum computers},\ \href
  {https://link.springer.com/article/10.1007/s11467-022-1249-z} {\bibfield
  {journal} {\bibinfo  {journal} {Front. Phys.}\ }\textbf {\bibinfo {volume}
  {18}},\ \bibinfo {pages} {21308} (\bibinfo {year} {2023})}\BibitemShut
  {NoStop}%
\bibitem [{\citenamefont {Jordan}\ \emph {et~al.}(2012)\citenamefont {Jordan},
  \citenamefont {Lee},\ and\ \citenamefont {Preskill}}]{jordan2012quantum}%
  \BibitemOpen
  \bibfield  {author} {\bibinfo {author} {\bibfnamefont {S.~P.}\ \bibnamefont
  {Jordan}}, \bibinfo {author} {\bibfnamefont {K.~S.}\ \bibnamefont {Lee}},\
  and\ \bibinfo {author} {\bibfnamefont {J.}~\bibnamefont {Preskill}},\
  }\bibinfo {title} {Quantum algorithms for quantum field theories},\ \href
  {https://doi.org/10.1126/science.1217069} {\bibfield  {journal} {\bibinfo
  {journal} {Science}\ }\textbf {\bibinfo {volume} {336}},\ \bibinfo {pages}
  {1130} (\bibinfo {year} {2012})}\BibitemShut {NoStop}%
\bibitem [{\citenamefont {Dalmonte}\ and\ \citenamefont
  {Montangero}(2016)}]{doi:10.1080/00107514.2016.1151199}%
  \BibitemOpen
  \bibfield  {author} {\bibinfo {author} {\bibfnamefont {M.}~\bibnamefont
  {Dalmonte}}\ and\ \bibinfo {author} {\bibfnamefont {S.}~\bibnamefont
  {Montangero}},\ }\bibinfo {title} {Lattice gauge theory simulations in the
  quantum information era},\ \href
  {https://doi.org/10.1080/00107514.2016.1151199} {\bibfield  {journal}
  {\bibinfo  {journal} {Contemp. Phys.}\ }\textbf {\bibinfo {volume} {57}},\
  \bibinfo {pages} {388} (\bibinfo {year} {2016})}\BibitemShut {NoStop}%
\bibitem [{\citenamefont {Banuls}\ \emph {et~al.}(2020)\citenamefont {Banuls},
  \citenamefont {Blatt}, \citenamefont {Catani}, \citenamefont {Celi},
  \citenamefont {Cirac}, \citenamefont {Dalmonte}, \citenamefont {Fallani},
  \citenamefont {Jansen}, \citenamefont {Lewenstein}, \citenamefont
  {Montangero} \emph {et~al.}}]{banuls2020simulating}%
  \BibitemOpen
  \bibfield  {author} {\bibinfo {author} {\bibfnamefont {M.~C.}\ \bibnamefont
  {Banuls}}, \bibinfo {author} {\bibfnamefont {R.}~\bibnamefont {Blatt}},
  \bibinfo {author} {\bibfnamefont {J.}~\bibnamefont {Catani}}, \bibinfo
  {author} {\bibfnamefont {A.}~\bibnamefont {Celi}}, \bibinfo {author}
  {\bibfnamefont {J.~I.}\ \bibnamefont {Cirac}}, \bibinfo {author}
  {\bibfnamefont {M.}~\bibnamefont {Dalmonte}}, \bibinfo {author}
  {\bibfnamefont {L.}~\bibnamefont {Fallani}}, \bibinfo {author} {\bibfnamefont
  {K.}~\bibnamefont {Jansen}}, \bibinfo {author} {\bibfnamefont
  {M.}~\bibnamefont {Lewenstein}}, \bibinfo {author} {\bibfnamefont
  {S.}~\bibnamefont {Montangero}}, \emph {et~al.},\ }\bibinfo {title}
  {Simulating lattice gauge theories within quantum technologies},\ \href
  {https://doi.org/10.1140/epjd/e2020-100571-8} {\bibfield  {journal} {\bibinfo
   {journal} {Eur. Phys. J. D}\ }\textbf {\bibinfo {volume} {74}},\ \bibinfo
  {pages} {1} (\bibinfo {year} {2020})}\BibitemShut {NoStop}%
\bibitem [{\citenamefont {Davoudi}\ \emph {et~al.}(2022)\citenamefont
  {Davoudi}, \citenamefont {Balantekin}, \citenamefont {Bhattacharya},
  \citenamefont {Carena}, \citenamefont {de~Jong}, \citenamefont {Draper},
  \citenamefont {El-Khadra}, \citenamefont {Gemelke}, \citenamefont {Hanada},
  \citenamefont {Kharzeev} \emph {et~al.}}]{davoudi2022quantum}%
  \BibitemOpen
  \bibfield  {author} {\bibinfo {author} {\bibfnamefont {C.~W.~B.}\
  \bibnamefont {Davoudi}}, \bibinfo {author} {\bibfnamefont {A.}~\bibnamefont
  {Balantekin}}, \bibinfo {author} {\bibfnamefont {T.}~\bibnamefont
  {Bhattacharya}}, \bibinfo {author} {\bibfnamefont {M.}~\bibnamefont
  {Carena}}, \bibinfo {author} {\bibfnamefont {W.~A.}\ \bibnamefont {de~Jong}},
  \bibinfo {author} {\bibfnamefont {P.}~\bibnamefont {Draper}}, \bibinfo
  {author} {\bibfnamefont {A.}~\bibnamefont {El-Khadra}}, \bibinfo {author}
  {\bibfnamefont {N.}~\bibnamefont {Gemelke}}, \bibinfo {author} {\bibfnamefont
  {M.}~\bibnamefont {Hanada}}, \bibinfo {author} {\bibfnamefont
  {D.}~\bibnamefont {Kharzeev}}, \emph {et~al.},\ }\bibinfo {title} {Quantum
  Simulation for High Energy Physics},\ \href
  {https://arxiv.org/abs/2204.03381} {\bibfield  {journal} {\bibinfo  {journal}
  {arXiv:2204.03381}\ } (\bibinfo {year} {2022})}\BibitemShut {NoStop}%
\bibitem [{\citenamefont {Bauer}\ \emph {et~al.}(2023)\citenamefont {Bauer},
  \citenamefont {Davoudi}, \citenamefont {Klco},\ and\ \citenamefont
  {Savage}}]{bauer2023quantum}%
  \BibitemOpen
  \bibfield  {author} {\bibinfo {author} {\bibfnamefont {C.~W.}\ \bibnamefont
  {Bauer}}, \bibinfo {author} {\bibfnamefont {Z.}~\bibnamefont {Davoudi}},
  \bibinfo {author} {\bibfnamefont {N.}~\bibnamefont {Klco}},\ and\ \bibinfo
  {author} {\bibfnamefont {M.~J.}\ \bibnamefont {Savage}},\ }\bibinfo {title}
  {Quantum simulation of fundamental particles and forces},\ \href
  {https://doi.org/10.1038/s42254-023-00599-8} {\bibfield  {journal} {\bibinfo
  {journal} {Nat. Rev. Phys.}\ }\textbf {\bibinfo {volume} {5}},\ \bibinfo
  {pages} {420} (\bibinfo {year} {2023})}\BibitemShut {NoStop}%
\bibitem [{\citenamefont {Halimeh}\ \emph {et~al.}(2024)\citenamefont
  {Halimeh}, \citenamefont {Hanada}, \citenamefont {Matsuura}, \citenamefont
  {Nori}, \citenamefont {Rinaldi},\ and\ \citenamefont
  {Sch{\"a}fer}}]{halimeh2024universal}%
  \BibitemOpen
  \bibfield  {author} {\bibinfo {author} {\bibfnamefont {J.~C.}\ \bibnamefont
  {Halimeh}}, \bibinfo {author} {\bibfnamefont {M.}~\bibnamefont {Hanada}},
  \bibinfo {author} {\bibfnamefont {S.}~\bibnamefont {Matsuura}}, \bibinfo
  {author} {\bibfnamefont {F.}~\bibnamefont {Nori}}, \bibinfo {author}
  {\bibfnamefont {E.}~\bibnamefont {Rinaldi}},\ and\ \bibinfo {author}
  {\bibfnamefont {A.}~\bibnamefont {Sch{\"a}fer}},\ }\bibinfo {title} {A
  universal framework for the quantum simulation of Yang-Mills theory},\ \href
  {https://arxiv.org/abs/2411.13161} {\bibfield  {journal} {\bibinfo  {journal}
  {arXiv:2411.13161}\ } (\bibinfo {year} {2024})}\BibitemShut {NoStop}%
\bibitem [{\citenamefont {Yang}\ \emph
  {et~al.}(2020{\natexlab{a}})\citenamefont {Yang}, \citenamefont {Sun},
  \citenamefont {Ott}, \citenamefont {Wang}, \citenamefont {Zache},
  \citenamefont {Halimeh}, \citenamefont {Yuan}, \citenamefont {Hauke},\ and\
  \citenamefont {Pan}}]{ISI:000591335100019}%
  \BibitemOpen
  \bibfield  {author} {\bibinfo {author} {\bibfnamefont {B.}~\bibnamefont
  {Yang}}, \bibinfo {author} {\bibfnamefont {H.}~\bibnamefont {Sun}}, \bibinfo
  {author} {\bibfnamefont {R.}~\bibnamefont {Ott}}, \bibinfo {author}
  {\bibfnamefont {H.-Y.}\ \bibnamefont {Wang}}, \bibinfo {author}
  {\bibfnamefont {T.~V.}\ \bibnamefont {Zache}}, \bibinfo {author}
  {\bibfnamefont {J.~C.}\ \bibnamefont {Halimeh}}, \bibinfo {author}
  {\bibfnamefont {Z.-S.}\ \bibnamefont {Yuan}}, \bibinfo {author}
  {\bibfnamefont {P.}~\bibnamefont {Hauke}},\ and\ \bibinfo {author}
  {\bibfnamefont {J.-W.}\ \bibnamefont {Pan}},\ }\bibinfo {title} {Observation
  of gauge invariance in a 71-site Bose-Hubbard quantum simulator},\ \href
  {https://doi.org/10.1038/s41586-020-2910-8} {\bibfield  {journal} {\bibinfo
  {journal} {Nature (London)}\ }\textbf {\bibinfo {volume} {587}},\ \bibinfo
  {pages} {392} (\bibinfo {year} {2020}{\natexlab{a}})}\BibitemShut {NoStop}%
\bibitem [{\citenamefont {Zhou}\ \emph {et~al.}(2022)\citenamefont {Zhou},
  \citenamefont {Su}, \citenamefont {Halimeh}, \citenamefont {Ott},
  \citenamefont {Sun}, \citenamefont {Hauke}, \citenamefont {Yang},
  \citenamefont {Yuan}, \citenamefont {Berges},\ and\ \citenamefont
  {Pan}}]{doi:10.1126/science.abl6277}%
  \BibitemOpen
  \bibfield  {author} {\bibinfo {author} {\bibfnamefont {Z.-Y.}\ \bibnamefont
  {Zhou}}, \bibinfo {author} {\bibfnamefont {G.-X.}\ \bibnamefont {Su}},
  \bibinfo {author} {\bibfnamefont {J.~C.}\ \bibnamefont {Halimeh}}, \bibinfo
  {author} {\bibfnamefont {R.}~\bibnamefont {Ott}}, \bibinfo {author}
  {\bibfnamefont {H.}~\bibnamefont {Sun}}, \bibinfo {author} {\bibfnamefont
  {P.}~\bibnamefont {Hauke}}, \bibinfo {author} {\bibfnamefont
  {B.}~\bibnamefont {Yang}}, \bibinfo {author} {\bibfnamefont {Z.-S.}\
  \bibnamefont {Yuan}}, \bibinfo {author} {\bibfnamefont {J.}~\bibnamefont
  {Berges}},\ and\ \bibinfo {author} {\bibfnamefont {J.-W.}\ \bibnamefont
  {Pan}},\ }\bibinfo {title} {Thermalization dynamics of a gauge theory on a
  quantum simulator},\ \href {https://doi.org/10.1126/science.abl6277}
  {\bibfield  {journal} {\bibinfo  {journal} {Science}\ }\textbf {\bibinfo
  {volume} {377}},\ \bibinfo {pages} {311} (\bibinfo {year}
  {2022})}\BibitemShut {NoStop}%
\bibitem [{\citenamefont {Zhang}\ \emph {et~al.}(2024)\citenamefont {Zhang},
  \citenamefont {Liu}, \citenamefont {Cheng}, \citenamefont {He}, \citenamefont
  {Wang}, \citenamefont {Wang}, \citenamefont {Zhu}, \citenamefont {Su},
  \citenamefont {Zhou}, \citenamefont {Zheng} \emph
  {et~al.}}]{zhang2024observation}%
  \BibitemOpen
  \bibfield  {author} {\bibinfo {author} {\bibfnamefont {W.-Y.}\ \bibnamefont
  {Zhang}}, \bibinfo {author} {\bibfnamefont {Y.}~\bibnamefont {Liu}}, \bibinfo
  {author} {\bibfnamefont {Y.}~\bibnamefont {Cheng}}, \bibinfo {author}
  {\bibfnamefont {M.-G.}\ \bibnamefont {He}}, \bibinfo {author} {\bibfnamefont
  {H.-Y.}\ \bibnamefont {Wang}}, \bibinfo {author} {\bibfnamefont {T.-Y.}\
  \bibnamefont {Wang}}, \bibinfo {author} {\bibfnamefont {Z.-H.}\ \bibnamefont
  {Zhu}}, \bibinfo {author} {\bibfnamefont {G.-X.}\ \bibnamefont {Su}},
  \bibinfo {author} {\bibfnamefont {Z.-Y.}\ \bibnamefont {Zhou}}, \bibinfo
  {author} {\bibfnamefont {Y.-G.}\ \bibnamefont {Zheng}}, \emph {et~al.},\
  }\bibinfo {title} {Observation of microscopic confinement dynamics by a
  tunable topological $\theta$-angle},\ \href
  {https://doi.org/10.1038/s41567-024-02702-x} {\bibfield  {journal} {\bibinfo
  {journal} {Nature Physics}\ }\textbf {\bibinfo {volume} {21}},\ \bibinfo
  {pages} {155} (\bibinfo {year} {2024})}\BibitemShut {NoStop}%
\bibitem [{\citenamefont {Gonz{\'a}lez-Cuadra}\ \emph
  {et~al.}(2025)\citenamefont {Gonz{\'a}lez-Cuadra}, \citenamefont {Hamdan},
  \citenamefont {Zache}, \citenamefont {Braverman}, \citenamefont
  {Kornja{\v{c}}a}, \citenamefont {Lukin}, \citenamefont {Cant{\'u}},
  \citenamefont {Liu}, \citenamefont {Wang}, \citenamefont {Keesling} \emph
  {et~al.}}]{gonzalez2025observation}%
  \BibitemOpen
  \bibfield  {author} {\bibinfo {author} {\bibfnamefont {D.}~\bibnamefont
  {Gonz{\'a}lez-Cuadra}}, \bibinfo {author} {\bibfnamefont {M.}~\bibnamefont
  {Hamdan}}, \bibinfo {author} {\bibfnamefont {T.~V.}\ \bibnamefont {Zache}},
  \bibinfo {author} {\bibfnamefont {B.}~\bibnamefont {Braverman}}, \bibinfo
  {author} {\bibfnamefont {M.}~\bibnamefont {Kornja{\v{c}}a}}, \bibinfo
  {author} {\bibfnamefont {A.}~\bibnamefont {Lukin}}, \bibinfo {author}
  {\bibfnamefont {S.~H.}\ \bibnamefont {Cant{\'u}}}, \bibinfo {author}
  {\bibfnamefont {F.}~\bibnamefont {Liu}}, \bibinfo {author} {\bibfnamefont
  {S.-T.}\ \bibnamefont {Wang}}, \bibinfo {author} {\bibfnamefont
  {A.}~\bibnamefont {Keesling}}, \emph {et~al.},\ }\bibinfo {title}
  {Observation of string breaking on a (2+ 1) D Rydberg quantum simulator},\
  \href {https://doi.org/10.1038/s41586-025-09051-6} {\bibfield  {journal}
  {\bibinfo  {journal} {Nature}\ ,\ \bibinfo {pages} {1}} (\bibinfo {year}
  {2025})}\BibitemShut {NoStop}%
\bibitem [{\citenamefont {Goerg}\ \emph {et~al.}(2019)\citenamefont {Goerg},
  \citenamefont {Sandholzer}, \citenamefont {Minguzzi}, \citenamefont
  {Desbuquois}, \citenamefont {Messer},\ and\ \citenamefont
  {Esslinger}}]{ISI:000494944200023}%
  \BibitemOpen
  \bibfield  {author} {\bibinfo {author} {\bibfnamefont {F.}~\bibnamefont
  {Goerg}}, \bibinfo {author} {\bibfnamefont {K.}~\bibnamefont {Sandholzer}},
  \bibinfo {author} {\bibfnamefont {J.}~\bibnamefont {Minguzzi}}, \bibinfo
  {author} {\bibfnamefont {R.}~\bibnamefont {Desbuquois}}, \bibinfo {author}
  {\bibfnamefont {M.}~\bibnamefont {Messer}},\ and\ \bibinfo {author}
  {\bibfnamefont {T.}~\bibnamefont {Esslinger}},\ }\bibinfo {title}
  {Realization of density-dependent Peierls phases to engineer quantized gauge
  fields coupled to ultracold matter},\ \href
  {https://doi.org/10.1038/s41567-019-0615-4} {\bibfield  {journal} {\bibinfo
  {journal} {Nat. Phys.}\ }\textbf {\bibinfo {volume} {15}},\ \bibinfo {pages}
  {1161} (\bibinfo {year} {2019})}\BibitemShut {NoStop}%
\bibitem [{\citenamefont {Schweizer}\ \emph {et~al.}(2019)\citenamefont
  {Schweizer}, \citenamefont {Grusdt}, \citenamefont {Berngruber},
  \citenamefont {Barbiero}, \citenamefont {Demler}, \citenamefont {Goldman},
  \citenamefont {Bloch},\ and\ \citenamefont
  {Aidelsburger}}]{ISI:000494944200024}%
  \BibitemOpen
  \bibfield  {author} {\bibinfo {author} {\bibfnamefont {C.}~\bibnamefont
  {Schweizer}}, \bibinfo {author} {\bibfnamefont {F.}~\bibnamefont {Grusdt}},
  \bibinfo {author} {\bibfnamefont {M.}~\bibnamefont {Berngruber}}, \bibinfo
  {author} {\bibfnamefont {L.}~\bibnamefont {Barbiero}}, \bibinfo {author}
  {\bibfnamefont {E.}~\bibnamefont {Demler}}, \bibinfo {author} {\bibfnamefont
  {N.}~\bibnamefont {Goldman}}, \bibinfo {author} {\bibfnamefont
  {I.}~\bibnamefont {Bloch}},\ and\ \bibinfo {author} {\bibfnamefont
  {M.}~\bibnamefont {Aidelsburger}},\ }\bibinfo {title} {Floquet approach to
  Z(2) lattice gauge theories with ultracold atoms in optical lattices},\ \href
  {https://doi.org/10.1038/s41567-019-0649-7} {\bibfield  {journal} {\bibinfo
  {journal} {Nat. Phys.}\ }\textbf {\bibinfo {volume} {15}},\ \bibinfo {pages}
  {1168} (\bibinfo {year} {2019})}\BibitemShut {NoStop}%
\bibitem [{\citenamefont {Wang}\ \emph {et~al.}(2022)\citenamefont {Wang},
  \citenamefont {Ge}, \citenamefont {Xiang}, \citenamefont {Song},
  \citenamefont {Huang}, \citenamefont {Song}, \citenamefont {Guo},
  \citenamefont {Su}, \citenamefont {Xu}, \citenamefont {Zheng},\ and\
  \citenamefont {Fan}}]{PhysRevResearch.4.L022060}%
  \BibitemOpen
  \bibfield  {author} {\bibinfo {author} {\bibfnamefont {Z.}~\bibnamefont
  {Wang}}, \bibinfo {author} {\bibfnamefont {Z.-Y.}\ \bibnamefont {Ge}},
  \bibinfo {author} {\bibfnamefont {Z.}~\bibnamefont {Xiang}}, \bibinfo
  {author} {\bibfnamefont {X.}~\bibnamefont {Song}}, \bibinfo {author}
  {\bibfnamefont {R.-Z.}\ \bibnamefont {Huang}}, \bibinfo {author}
  {\bibfnamefont {P.}~\bibnamefont {Song}}, \bibinfo {author} {\bibfnamefont
  {X.-Y.}\ \bibnamefont {Guo}}, \bibinfo {author} {\bibfnamefont
  {L.}~\bibnamefont {Su}}, \bibinfo {author} {\bibfnamefont {K.}~\bibnamefont
  {Xu}}, \bibinfo {author} {\bibfnamefont {D.}~\bibnamefont {Zheng}},\ and\
  \bibinfo {author} {\bibfnamefont {H.}~\bibnamefont {Fan}},\ }\bibinfo {title}
  {Observation of emergent ${\mathbb{Z}}_{2}$ gauge invariance in a
  superconducting circuit},\ \href
  {https://doi.org/10.1103/PhysRevResearch.4.L022060} {\bibfield  {journal}
  {\bibinfo  {journal} {Phys. Rev. Res.}\ }\textbf {\bibinfo {volume} {4}},\
  \bibinfo {pages} {L022060} (\bibinfo {year} {2022})}\BibitemShut {NoStop}%
\bibitem [{\citenamefont {Mildenberger}\ \emph {et~al.}(2025)\citenamefont
  {Mildenberger}, \citenamefont {Mruczkiewicz}, \citenamefont {Halimeh},
  \citenamefont {Jiang},\ and\ \citenamefont
  {Hauke}}]{mildenberger2025confinement}%
  \BibitemOpen
  \bibfield  {author} {\bibinfo {author} {\bibfnamefont {J.}~\bibnamefont
  {Mildenberger}}, \bibinfo {author} {\bibfnamefont {W.}~\bibnamefont
  {Mruczkiewicz}}, \bibinfo {author} {\bibfnamefont {J.~C.}\ \bibnamefont
  {Halimeh}}, \bibinfo {author} {\bibfnamefont {Z.}~\bibnamefont {Jiang}},\
  and\ \bibinfo {author} {\bibfnamefont {P.}~\bibnamefont {Hauke}},\ }\bibinfo
  {title} {Confinement in a $Z_2$ lattice gauge theory on a quantum computer},\
  \href {https://www.nature.com/articles/s41567-024-02723-6} {\bibfield
  {journal} {\bibinfo  {journal} {Nat. Phys.}\ }\textbf {\bibinfo {volume}
  {21}},\ \bibinfo {pages} {312} (\bibinfo {year} {2025})}\BibitemShut
  {NoStop}%
\bibitem [{\citenamefont {De}\ \emph {et~al.}(2024)\citenamefont {De},
  \citenamefont {Lerose}, \citenamefont {Luo}, \citenamefont {Surace},
  \citenamefont {Schuckert}, \citenamefont {Bennewitz}, \citenamefont {Ware},
  \citenamefont {Morong}, \citenamefont {Collins}, \citenamefont {Davoudi}
  \emph {et~al.}}]{de2024observation}%
  \BibitemOpen
  \bibfield  {author} {\bibinfo {author} {\bibfnamefont {A.}~\bibnamefont
  {De}}, \bibinfo {author} {\bibfnamefont {A.}~\bibnamefont {Lerose}}, \bibinfo
  {author} {\bibfnamefont {D.}~\bibnamefont {Luo}}, \bibinfo {author}
  {\bibfnamefont {F.~M.}\ \bibnamefont {Surace}}, \bibinfo {author}
  {\bibfnamefont {A.}~\bibnamefont {Schuckert}}, \bibinfo {author}
  {\bibfnamefont {E.~R.}\ \bibnamefont {Bennewitz}}, \bibinfo {author}
  {\bibfnamefont {B.}~\bibnamefont {Ware}}, \bibinfo {author} {\bibfnamefont
  {W.}~\bibnamefont {Morong}}, \bibinfo {author} {\bibfnamefont {K.~S.}\
  \bibnamefont {Collins}}, \bibinfo {author} {\bibfnamefont {Z.}~\bibnamefont
  {Davoudi}}, \emph {et~al.},\ }\bibinfo {title} {Observation of
  string-breaking dynamics in a quantum simulator},\ \href
  {https://arxiv.org/abs/2410.13815} {\bibfield  {journal} {\bibinfo  {journal}
  {arXiv:2410.13815}\ } (\bibinfo {year} {2024})}\BibitemShut {NoStop}%
\bibitem [{\citenamefont {Cochran}\ \emph {et~al.}(2025)\citenamefont
  {Cochran}, \citenamefont {Jobst}, \citenamefont {Rosenberg}, \citenamefont
  {Lensky}, \citenamefont {Gyawali}, \citenamefont {Eassa}, \citenamefont
  {Will}, \citenamefont {Szasz}, \citenamefont {Abanin}, \citenamefont
  {Acharya} \emph {et~al.}}]{cochran2025visualizing}%
  \BibitemOpen
  \bibfield  {author} {\bibinfo {author} {\bibfnamefont {T.~A.}\ \bibnamefont
  {Cochran}}, \bibinfo {author} {\bibfnamefont {B.}~\bibnamefont {Jobst}},
  \bibinfo {author} {\bibfnamefont {E.}~\bibnamefont {Rosenberg}}, \bibinfo
  {author} {\bibfnamefont {Y.~D.}\ \bibnamefont {Lensky}}, \bibinfo {author}
  {\bibfnamefont {G.}~\bibnamefont {Gyawali}}, \bibinfo {author} {\bibfnamefont
  {N.}~\bibnamefont {Eassa}}, \bibinfo {author} {\bibfnamefont
  {M.}~\bibnamefont {Will}}, \bibinfo {author} {\bibfnamefont {A.}~\bibnamefont
  {Szasz}}, \bibinfo {author} {\bibfnamefont {D.}~\bibnamefont {Abanin}},
  \bibinfo {author} {\bibfnamefont {R.}~\bibnamefont {Acharya}}, \emph
  {et~al.},\ }\bibinfo {title} {Visualizing dynamics of charges and strings in
  (2+1)D lattice gauge theories},\ \href
  {https://doi.org/10.1038/s41586-025-08999-9} {\bibfield  {journal} {\bibinfo
  {journal} {Nature}\ }\textbf {\bibinfo {volume} {315}},\ \bibinfo {pages}
  {642} (\bibinfo {year} {2025})}\BibitemShut {NoStop}%
\bibitem [{\citenamefont {Kormos}\ \emph {et~al.}(2017)\citenamefont {Kormos},
  \citenamefont {Collura}, \citenamefont {Takcs},\ and\ \citenamefont
  {Calabrese}}]{ISI:000395814000014}%
  \BibitemOpen
  \bibfield  {author} {\bibinfo {author} {\bibfnamefont {M.}~\bibnamefont
  {Kormos}}, \bibinfo {author} {\bibfnamefont {M.}~\bibnamefont {Collura}},
  \bibinfo {author} {\bibfnamefont {G.}~\bibnamefont {Takcs}},\ and\ \bibinfo
  {author} {\bibfnamefont {P.}~\bibnamefont {Calabrese}},\ }\bibinfo {title}
  {Real-time confinement following a quantum quench to a non-integrable
  model},\ \href {https://doi.org/10.1038/NPHYS3934} {\bibfield  {journal}
  {\bibinfo  {journal} {Nat. Phys.}\ }\textbf {\bibinfo {volume} {13}},\
  \bibinfo {pages} {246} (\bibinfo {year} {2017})}\BibitemShut {NoStop}%
\bibitem [{\citenamefont {Hebenstreit}\ \emph {et~al.}(2013)\citenamefont
  {Hebenstreit}, \citenamefont {Berges},\ and\ \citenamefont
  {Gelfand}}]{PhysRevLett.111.201601}%
  \BibitemOpen
  \bibfield  {author} {\bibinfo {author} {\bibfnamefont {F.}~\bibnamefont
  {Hebenstreit}}, \bibinfo {author} {\bibfnamefont {J.}~\bibnamefont
  {Berges}},\ and\ \bibinfo {author} {\bibfnamefont {D.}~\bibnamefont
  {Gelfand}},\ }\bibinfo {title} {Real-time dynamics of string breaking},\
  \href {https://doi.org/10.1103/PhysRevLett.111.201601} {\bibfield  {journal}
  {\bibinfo  {journal} {Phys. Rev. Lett.}\ }\textbf {\bibinfo {volume} {111}},\
  \bibinfo {pages} {201601} (\bibinfo {year} {2013})}\BibitemShut {NoStop}%
\bibitem [{\citenamefont {Jurcevic}\ \emph {et~al.}(2015)\citenamefont
  {Jurcevic}, \citenamefont {Hauke}, \citenamefont {Maier}, \citenamefont
  {Hempel}, \citenamefont {Lanyon}, \citenamefont {Blatt},\ and\ \citenamefont
  {Roos}}]{PhysRevLett.115.100501}%
  \BibitemOpen
  \bibfield  {author} {\bibinfo {author} {\bibfnamefont {P.}~\bibnamefont
  {Jurcevic}}, \bibinfo {author} {\bibfnamefont {P.}~\bibnamefont {Hauke}},
  \bibinfo {author} {\bibfnamefont {C.}~\bibnamefont {Maier}}, \bibinfo
  {author} {\bibfnamefont {C.}~\bibnamefont {Hempel}}, \bibinfo {author}
  {\bibfnamefont {B.~P.}\ \bibnamefont {Lanyon}}, \bibinfo {author}
  {\bibfnamefont {R.}~\bibnamefont {Blatt}},\ and\ \bibinfo {author}
  {\bibfnamefont {C.~F.}\ \bibnamefont {Roos}},\ }\bibinfo {title}
  {Spectroscopy of Interacting Quasiparticles in Trapped Ions},\ \href
  {https://doi.org/10.1103/PhysRevLett.115.100501} {\bibfield  {journal}
  {\bibinfo  {journal} {Phys. Rev. Lett.}\ }\textbf {\bibinfo {volume} {115}},\
  \bibinfo {pages} {100501} (\bibinfo {year} {2015})}\BibitemShut {NoStop}%
\bibitem [{\citenamefont {Kranzl}\ \emph {et~al.}(2023)\citenamefont {Kranzl},
  \citenamefont {Birnkammer}, \citenamefont {Joshi}, \citenamefont
  {Bastianello}, \citenamefont {Blatt}, \citenamefont {Knap},\ and\
  \citenamefont {Roos}}]{PhysRevX.13.031017}%
  \BibitemOpen
  \bibfield  {author} {\bibinfo {author} {\bibfnamefont {F.}~\bibnamefont
  {Kranzl}}, \bibinfo {author} {\bibfnamefont {S.}~\bibnamefont {Birnkammer}},
  \bibinfo {author} {\bibfnamefont {M.~K.}\ \bibnamefont {Joshi}}, \bibinfo
  {author} {\bibfnamefont {A.}~\bibnamefont {Bastianello}}, \bibinfo {author}
  {\bibfnamefont {R.}~\bibnamefont {Blatt}}, \bibinfo {author} {\bibfnamefont
  {M.}~\bibnamefont {Knap}},\ and\ \bibinfo {author} {\bibfnamefont {C.~F.}\
  \bibnamefont {Roos}},\ }\bibinfo {title} {Observation of Magnon Bound States
  in the Long-Range, Anisotropic Heisenberg Model},\ \href
  {https://doi.org/10.1103/PhysRevX.13.031017} {\bibfield  {journal} {\bibinfo
  {journal} {Phys. Rev. X}\ }\textbf {\bibinfo {volume} {13}},\ \bibinfo
  {pages} {031017} (\bibinfo {year} {2023})}\BibitemShut {NoStop}%
\bibitem [{\citenamefont {Ellis}\ \emph {et~al.}(2003)\citenamefont {Ellis},
  \citenamefont {Stirling},\ and\ \citenamefont {Webber}}]{ellis2003qcd}%
  \BibitemOpen
  \bibfield  {author} {\bibinfo {author} {\bibfnamefont {R.~K.}\ \bibnamefont
  {Ellis}}, \bibinfo {author} {\bibfnamefont {W.~J.}\ \bibnamefont
  {Stirling}},\ and\ \bibinfo {author} {\bibfnamefont {B.~R.}\ \bibnamefont
  {Webber}},\ }\href@noop {} {\emph {\bibinfo {title} {QCD and collider
  physics}}},\ \bibinfo {number} {8}\ (\bibinfo  {publisher} {Cambridge
  university press},\ \bibinfo {year} {2003})\BibitemShut {NoStop}%
\bibitem [{\citenamefont {Achenbach}\ \emph {et~al.}(2024)\citenamefont
  {Achenbach}, \citenamefont {Adhikari}, \citenamefont {Afanasev} \emph
  {et~al.}}]{ACHENBACH2024122874}%
  \BibitemOpen
  \bibfield  {author} {\bibinfo {author} {\bibfnamefont {P.}~\bibnamefont
  {Achenbach}}, \bibinfo {author} {\bibfnamefont {D.}~\bibnamefont {Adhikari}},
  \bibinfo {author} {\bibfnamefont {A.}~\bibnamefont {Afanasev}}, \emph
  {et~al.},\ }\bibinfo {title} {The present and future of QCD},\ \href
  {https://doi.org/https://doi.org/10.1016/j.nuclphysa.2024.122874} {\bibfield
  {journal} {\bibinfo  {journal} {Nucl. Phys. A}\ }\textbf {\bibinfo {volume}
  {1047}},\ \bibinfo {pages} {122874} (\bibinfo {year} {2024})}\BibitemShut
  {NoStop}%
\bibitem [{\citenamefont {Borla}\ \emph {et~al.}(2020)\citenamefont {Borla},
  \citenamefont {Verresen}, \citenamefont {Grusdt},\ and\ \citenamefont
  {Moroz}}]{PhysRevLett.124.120503}%
  \BibitemOpen
  \bibfield  {author} {\bibinfo {author} {\bibfnamefont {U.}~\bibnamefont
  {Borla}}, \bibinfo {author} {\bibfnamefont {R.}~\bibnamefont {Verresen}},
  \bibinfo {author} {\bibfnamefont {F.}~\bibnamefont {Grusdt}},\ and\ \bibinfo
  {author} {\bibfnamefont {S.}~\bibnamefont {Moroz}},\ }\bibinfo {title}
  {Confined phases of one-dimensional spinless Fermions coupled to ${Z}_{2}$
  gauge theory},\ \href {https://doi.org/10.1103/PhysRevLett.124.120503}
  {\bibfield  {journal} {\bibinfo  {journal} {Phys. Rev. Lett.}\ }\textbf
  {\bibinfo {volume} {124}},\ \bibinfo {pages} {120503} (\bibinfo {year}
  {2020})}\BibitemShut {NoStop}%
\bibitem [{\citenamefont {Liu}\ \emph {et~al.}(2019)\citenamefont {Liu},
  \citenamefont {Lundgren}, \citenamefont {Titum}, \citenamefont {Pagano},
  \citenamefont {Zhang}, \citenamefont {Monroe},\ and\ \citenamefont
  {Gorshkov}}]{PhysRevLett.122.150601}%
  \BibitemOpen
  \bibfield  {author} {\bibinfo {author} {\bibfnamefont {F.}~\bibnamefont
  {Liu}}, \bibinfo {author} {\bibfnamefont {R.}~\bibnamefont {Lundgren}},
  \bibinfo {author} {\bibfnamefont {P.}~\bibnamefont {Titum}}, \bibinfo
  {author} {\bibfnamefont {G.}~\bibnamefont {Pagano}}, \bibinfo {author}
  {\bibfnamefont {J.}~\bibnamefont {Zhang}}, \bibinfo {author} {\bibfnamefont
  {C.}~\bibnamefont {Monroe}},\ and\ \bibinfo {author} {\bibfnamefont {A.~V.}\
  \bibnamefont {Gorshkov}},\ }\bibinfo {title} {Confined Quasiparticle Dynamics
  in Long-Range Interacting Quantum Spin Chains},\ \href
  {https://doi.org/10.1103/PhysRevLett.122.150601} {\bibfield  {journal}
  {\bibinfo  {journal} {Phys. Rev. Lett.}\ }\textbf {\bibinfo {volume} {122}},\
  \bibinfo {pages} {150601} (\bibinfo {year} {2019})}\BibitemShut {NoStop}%
\bibitem [{\citenamefont {Surace}\ and\ \citenamefont
  {Lerose}(2021)}]{surace2021scattering}%
  \BibitemOpen
  \bibfield  {author} {\bibinfo {author} {\bibfnamefont {F.~M.}\ \bibnamefont
  {Surace}}\ and\ \bibinfo {author} {\bibfnamefont {A.}~\bibnamefont
  {Lerose}},\ }\bibinfo {title} {Scattering of mesons in quantum simulators},\
  \href {https://iopscience.iop.org/article/10.1088/1367-2630/abfc40/meta}
  {\bibfield  {journal} {\bibinfo  {journal} {New J. Phys.}\ }\textbf {\bibinfo
  {volume} {23}},\ \bibinfo {pages} {062001} (\bibinfo {year}
  {2021})}\BibitemShut {NoStop}%
\bibitem [{\citenamefont {Vovrosh}\ \emph {et~al.}(2022)\citenamefont
  {Vovrosh}, \citenamefont {Mukherjee}, \citenamefont {Bastianello},\ and\
  \citenamefont {Knolle}}]{PRXQuantum.3.040309}%
  \BibitemOpen
  \bibfield  {author} {\bibinfo {author} {\bibfnamefont {J.}~\bibnamefont
  {Vovrosh}}, \bibinfo {author} {\bibfnamefont {R.}~\bibnamefont {Mukherjee}},
  \bibinfo {author} {\bibfnamefont {A.}~\bibnamefont {Bastianello}},\ and\
  \bibinfo {author} {\bibfnamefont {J.}~\bibnamefont {Knolle}},\ }\bibinfo
  {title} {Dynamical Hadron Formation in Long-Range Interacting Quantum Spin
  Chains},\ \href {https://doi.org/10.1103/PRXQuantum.3.040309} {\bibfield
  {journal} {\bibinfo  {journal} {PRX Quantum}\ }\textbf {\bibinfo {volume}
  {3}},\ \bibinfo {pages} {040309} (\bibinfo {year} {2022})}\BibitemShut
  {NoStop}%
\bibitem [{\citenamefont {Karpov}\ \emph {et~al.}(2022)\citenamefont {Karpov},
  \citenamefont {Zhu}, \citenamefont {Heller},\ and\ \citenamefont
  {Heyl}}]{PhysRevResearch.4.L032001}%
  \BibitemOpen
  \bibfield  {author} {\bibinfo {author} {\bibfnamefont {P.~I.}\ \bibnamefont
  {Karpov}}, \bibinfo {author} {\bibfnamefont {G.-Y.}\ \bibnamefont {Zhu}},
  \bibinfo {author} {\bibfnamefont {M.~P.}\ \bibnamefont {Heller}},\ and\
  \bibinfo {author} {\bibfnamefont {M.}~\bibnamefont {Heyl}},\ }\bibinfo
  {title} {Spatiotemporal dynamics of particle collisions in quantum spin
  chains},\ \href {https://doi.org/10.1103/PhysRevResearch.4.L032001}
  {\bibfield  {journal} {\bibinfo  {journal} {Phys. Rev. Res.}\ }\textbf
  {\bibinfo {volume} {4}},\ \bibinfo {pages} {L032001} (\bibinfo {year}
  {2022})}\BibitemShut {NoStop}%
\bibitem [{\citenamefont {Turco}\ \emph {et~al.}(2024)\citenamefont {Turco},
  \citenamefont {Quinta}, \citenamefont {Seixas},\ and\ \citenamefont
  {Omar}}]{PRXQuantum.5.020311}%
  \BibitemOpen
  \bibfield  {author} {\bibinfo {author} {\bibfnamefont {M.}~\bibnamefont
  {Turco}}, \bibinfo {author} {\bibfnamefont {G.}~\bibnamefont {Quinta}},
  \bibinfo {author} {\bibfnamefont {J.}~\bibnamefont {Seixas}},\ and\ \bibinfo
  {author} {\bibfnamefont {Y.}~\bibnamefont {Omar}},\ }\bibinfo {title}
  {Quantum Simulation of Bound State Scattering},\ \href
  {https://doi.org/10.1103/PRXQuantum.5.020311} {\bibfield  {journal} {\bibinfo
   {journal} {PRX Quantum}\ }\textbf {\bibinfo {volume} {5}},\ \bibinfo {pages}
  {020311} (\bibinfo {year} {2024})}\BibitemShut {NoStop}%
\bibitem [{\citenamefont {Kreshchuk}\ \emph {et~al.}(2023)\citenamefont
  {Kreshchuk}, \citenamefont {Vary},\ and\ \citenamefont
  {Love}}]{kreshchuk2023simulating}%
  \BibitemOpen
  \bibfield  {author} {\bibinfo {author} {\bibfnamefont {M.}~\bibnamefont
  {Kreshchuk}}, \bibinfo {author} {\bibfnamefont {J.~P.}\ \bibnamefont
  {Vary}},\ and\ \bibinfo {author} {\bibfnamefont {P.~J.}\ \bibnamefont
  {Love}},\ }\bibinfo {title} {Simulating scattering of composite particles},\
  \href {https://arxiv.org/abs/2310.13742} {\bibfield  {journal} {\bibinfo
  {journal} {arXiv:2310.13742}\ } (\bibinfo {year} {2023})}\BibitemShut
  {NoStop}%
\bibitem [{\citenamefont {Chai}\ \emph {et~al.}(2025)\citenamefont {Chai},
  \citenamefont {Crippa}, \citenamefont {Jansen}, \citenamefont {K{\"{u}}hn},
  \citenamefont {Pascuzzi}, \citenamefont {Tacchino},\ and\ \citenamefont
  {Tavernelli}}]{Chai2025fermionicwavepacket}%
  \BibitemOpen
  \bibfield  {author} {\bibinfo {author} {\bibfnamefont {Y.}~\bibnamefont
  {Chai}}, \bibinfo {author} {\bibfnamefont {A.}~\bibnamefont {Crippa}},
  \bibinfo {author} {\bibfnamefont {K.}~\bibnamefont {Jansen}}, \bibinfo
  {author} {\bibfnamefont {S.}~\bibnamefont {K{\"{u}}hn}}, \bibinfo {author}
  {\bibfnamefont {V.~R.}\ \bibnamefont {Pascuzzi}}, \bibinfo {author}
  {\bibfnamefont {F.}~\bibnamefont {Tacchino}},\ and\ \bibinfo {author}
  {\bibfnamefont {I.}~\bibnamefont {Tavernelli}},\ }\bibinfo {title} {Fermionic
  wave packet scattering: a quantum computing approach},\ \href
  {https://doi.org/10.22331/q-2025-02-19-1638} {\bibfield  {journal} {\bibinfo
  {journal} {{Quantum}}\ }\textbf {\bibinfo {volume} {9}},\ \bibinfo {pages}
  {1638} (\bibinfo {year} {2025})}\BibitemShut {NoStop}%
\bibitem [{\citenamefont {Farrell}\ \emph {et~al.}(2024)\citenamefont
  {Farrell}, \citenamefont {Illa}, \citenamefont {Ciavarella},\ and\
  \citenamefont {Savage}}]{PhysRevD.109.114510}%
  \BibitemOpen
  \bibfield  {author} {\bibinfo {author} {\bibfnamefont {R.~C.}\ \bibnamefont
  {Farrell}}, \bibinfo {author} {\bibfnamefont {M.}~\bibnamefont {Illa}},
  \bibinfo {author} {\bibfnamefont {A.~N.}\ \bibnamefont {Ciavarella}},\ and\
  \bibinfo {author} {\bibfnamefont {M.~J.}\ \bibnamefont {Savage}},\ }\bibinfo
  {title} {Quantum simulations of hadron dynamics in the Schwinger model using
  112 qubits},\ \href {https://doi.org/10.1103/PhysRevD.109.114510} {\bibfield
  {journal} {\bibinfo  {journal} {Phys. Rev. D}\ }\textbf {\bibinfo {volume}
  {109}},\ \bibinfo {pages} {114510} (\bibinfo {year} {2024})}\BibitemShut
  {NoStop}%
\bibitem [{\citenamefont {Davoudi}\ \emph {et~al.}(2024)\citenamefont
  {Davoudi}, \citenamefont {Hsieh},\ and\ \citenamefont
  {Kadam}}]{Davoudi2024scatteringwave}%
  \BibitemOpen
  \bibfield  {author} {\bibinfo {author} {\bibfnamefont {Z.}~\bibnamefont
  {Davoudi}}, \bibinfo {author} {\bibfnamefont {C.-C.}\ \bibnamefont {Hsieh}},\
  and\ \bibinfo {author} {\bibfnamefont {S.~V.}\ \bibnamefont {Kadam}},\
  }\bibinfo {title} {Scattering wave packets of hadrons in gauge theories:
  {P}reparation on a quantum computer},\ \href
  {https://doi.org/10.22331/q-2024-11-11-1520} {\bibfield  {journal} {\bibinfo
  {journal} {{Quantum}}\ }\textbf {\bibinfo {volume} {8}},\ \bibinfo {pages}
  {1520} (\bibinfo {year} {2024})}\BibitemShut {NoStop}%
\bibitem [{\citenamefont {Su}\ \emph {et~al.}(2024)\citenamefont {Su},
  \citenamefont {Osborne},\ and\ \citenamefont
  {Halimeh}}]{PRXQuantum.5.040310}%
  \BibitemOpen
  \bibfield  {author} {\bibinfo {author} {\bibfnamefont {G.-X.}\ \bibnamefont
  {Su}}, \bibinfo {author} {\bibfnamefont {J.~J.}\ \bibnamefont {Osborne}},\
  and\ \bibinfo {author} {\bibfnamefont {J.~C.}\ \bibnamefont {Halimeh}},\
  }\bibinfo {title} {Cold-Atom Particle Collider},\ \href
  {https://doi.org/10.1103/PRXQuantum.5.040310} {\bibfield  {journal} {\bibinfo
   {journal} {PRX Quantum}\ }\textbf {\bibinfo {volume} {5}},\ \bibinfo {pages}
  {040310} (\bibinfo {year} {2024})}\BibitemShut {NoStop}%
\bibitem [{\citenamefont {Bennewitz}\ \emph {et~al.}(2025)\citenamefont
  {Bennewitz}, \citenamefont {Ware}, \citenamefont {Schuckert}, \citenamefont
  {Lerose}, \citenamefont {Surace}, \citenamefont {Belyansky}, \citenamefont
  {Morong}, \citenamefont {Luo}, \citenamefont {De}, \citenamefont {Collins},
  \citenamefont {Katz}, \citenamefont {Monroe}, \citenamefont {Davoudi},\ and\
  \citenamefont {Gorshkov}}]{Bennewitz2025simulatingmeson}%
  \BibitemOpen
  \bibfield  {author} {\bibinfo {author} {\bibfnamefont {E.~R.}\ \bibnamefont
  {Bennewitz}}, \bibinfo {author} {\bibfnamefont {B.}~\bibnamefont {Ware}},
  \bibinfo {author} {\bibfnamefont {A.}~\bibnamefont {Schuckert}}, \bibinfo
  {author} {\bibfnamefont {A.}~\bibnamefont {Lerose}}, \bibinfo {author}
  {\bibfnamefont {F.~M.}\ \bibnamefont {Surace}}, \bibinfo {author}
  {\bibfnamefont {R.}~\bibnamefont {Belyansky}}, \bibinfo {author}
  {\bibfnamefont {W.}~\bibnamefont {Morong}}, \bibinfo {author} {\bibfnamefont
  {D.}~\bibnamefont {Luo}}, \bibinfo {author} {\bibfnamefont {A.}~\bibnamefont
  {De}}, \bibinfo {author} {\bibfnamefont {K.~S.}\ \bibnamefont {Collins}},
  \bibinfo {author} {\bibfnamefont {O.}~\bibnamefont {Katz}}, \bibinfo {author}
  {\bibfnamefont {C.}~\bibnamefont {Monroe}}, \bibinfo {author} {\bibfnamefont
  {Z.}~\bibnamefont {Davoudi}},\ and\ \bibinfo {author} {\bibfnamefont {A.~V.}\
  \bibnamefont {Gorshkov}},\ }\bibinfo {title} {Simulating {M}eson {S}cattering
  on {S}pin {Q}uantum {S}imulators},\ \href
  {https://doi.org/10.22331/q-2025-06-17-1773} {\bibfield  {journal} {\bibinfo
  {journal} {{Quantum}}\ }\textbf {\bibinfo {volume} {9}},\ \bibinfo {pages}
  {1773} (\bibinfo {year} {2025})}\BibitemShut {NoStop}%
\bibitem [{\citenamefont {Schuhmacher}\ \emph {et~al.}(2025)\citenamefont
  {Schuhmacher}, \citenamefont {Su}, \citenamefont {Osborne}, \citenamefont
  {Gandon}, \citenamefont {Halimeh},\ and\ \citenamefont
  {Tavernelli}}]{schuhmacher2025observation}%
  \BibitemOpen
  \bibfield  {author} {\bibinfo {author} {\bibfnamefont {J.}~\bibnamefont
  {Schuhmacher}}, \bibinfo {author} {\bibfnamefont {G.-X.}\ \bibnamefont {Su}},
  \bibinfo {author} {\bibfnamefont {J.~J.}\ \bibnamefont {Osborne}}, \bibinfo
  {author} {\bibfnamefont {A.}~\bibnamefont {Gandon}}, \bibinfo {author}
  {\bibfnamefont {J.~C.}\ \bibnamefont {Halimeh}},\ and\ \bibinfo {author}
  {\bibfnamefont {I.}~\bibnamefont {Tavernelli}},\ }\bibinfo {title}
  {Observation of hadron scattering in a lattice gauge theory on a quantum
  computer},\ \href {https://arxiv.org/abs/2505.20387} {\bibfield  {journal}
  {\bibinfo  {journal} {arXiv:2505.20387}\ } (\bibinfo {year}
  {2025})}\BibitemShut {NoStop}%
\bibitem [{\citenamefont {Joshi}\ \emph {et~al.}(2025)\citenamefont {Joshi},
  \citenamefont {Louw}, \citenamefont {Meth}, \citenamefont {Osborne},
  \citenamefont {Mato}, \citenamefont {Su}, \citenamefont {Ringbauer},\ and\
  \citenamefont {Halimeh}}]{joshi2025probing}%
  \BibitemOpen
  \bibfield  {author} {\bibinfo {author} {\bibfnamefont {R.}~\bibnamefont
  {Joshi}}, \bibinfo {author} {\bibfnamefont {J.~C.}\ \bibnamefont {Louw}},
  \bibinfo {author} {\bibfnamefont {M.}~\bibnamefont {Meth}}, \bibinfo {author}
  {\bibfnamefont {J.~J.}\ \bibnamefont {Osborne}}, \bibinfo {author}
  {\bibfnamefont {K.}~\bibnamefont {Mato}}, \bibinfo {author} {\bibfnamefont
  {G.-X.}\ \bibnamefont {Su}}, \bibinfo {author} {\bibfnamefont
  {M.}~\bibnamefont {Ringbauer}},\ and\ \bibinfo {author} {\bibfnamefont
  {J.~C.}\ \bibnamefont {Halimeh}},\ }\bibinfo {title} {Probing Hadron
  Scattering in Lattice Gauge Theories on Qudit Quantum Computers},\ \href
  {https://arxiv.org/abs/2507.12614} {\bibfield  {journal} {\bibinfo  {journal}
  {arXiv:2507.12614}\ } (\bibinfo {year} {2025})}\BibitemShut {NoStop}%
\bibitem [{SM()}]{SM}%
  \BibitemOpen
  \href@noop {} {\bibinfo {title} {See supplemental material}}\BibitemShut
  {NoStop}%
\bibitem [{\citenamefont {Koch}\ \emph {et~al.}(2007)\citenamefont {Koch},
  \citenamefont {Yu}, \citenamefont {Gambetta}, \citenamefont {Houck},
  \citenamefont {Schuster}, \citenamefont {Majer}, \citenamefont {Blais},
  \citenamefont {Devoret}, \citenamefont {Girvin},\ and\ \citenamefont
  {Schoelkopf}}]{koch2007charge}%
  \BibitemOpen
  \bibfield  {author} {\bibinfo {author} {\bibfnamefont {J.}~\bibnamefont
  {Koch}}, \bibinfo {author} {\bibfnamefont {T.~M.}\ \bibnamefont {Yu}},
  \bibinfo {author} {\bibfnamefont {J.}~\bibnamefont {Gambetta}}, \bibinfo
  {author} {\bibfnamefont {A.~A.}\ \bibnamefont {Houck}}, \bibinfo {author}
  {\bibfnamefont {D.~I.}\ \bibnamefont {Schuster}}, \bibinfo {author}
  {\bibfnamefont {J.}~\bibnamefont {Majer}}, \bibinfo {author} {\bibfnamefont
  {A.}~\bibnamefont {Blais}}, \bibinfo {author} {\bibfnamefont {M.~H.}\
  \bibnamefont {Devoret}}, \bibinfo {author} {\bibfnamefont {S.~M.}\
  \bibnamefont {Girvin}},\ and\ \bibinfo {author} {\bibfnamefont {R.~J.}\
  \bibnamefont {Schoelkopf}},\ }\bibinfo {title} {Charge-insensitive qubit
  design derived from the {C}ooper pair box},\ \href
  {https://journals.aps.org/pra/abstract/10.1103/PhysRevA.76.042319} {\bibfield
   {journal} {\bibinfo  {journal} {Phys. Rev. A}\ }\textbf {\bibinfo {volume}
  {76}},\ \bibinfo {pages} {042319} (\bibinfo {year} {2007})}\BibitemShut
  {NoStop}%
\bibitem [{\citenamefont {Kjaergaard}\ \emph {et~al.}(2020)\citenamefont
  {Kjaergaard}, \citenamefont {Schwartz}, \citenamefont {Braumüller},
  \citenamefont {Krantz}, \citenamefont {Wang}, \citenamefont {Gustavsson},\
  and\ \citenamefont {Oliver}}]{annurev-conmatphys-031119-050605}%
  \BibitemOpen
  \bibfield  {author} {\bibinfo {author} {\bibfnamefont {M.}~\bibnamefont
  {Kjaergaard}}, \bibinfo {author} {\bibfnamefont {M.~E.}\ \bibnamefont
  {Schwartz}}, \bibinfo {author} {\bibfnamefont {J.}~\bibnamefont
  {Braumüller}}, \bibinfo {author} {\bibfnamefont {P.}~\bibnamefont {Krantz}},
  \bibinfo {author} {\bibfnamefont {J.~I.-J.}\ \bibnamefont {Wang}}, \bibinfo
  {author} {\bibfnamefont {S.}~\bibnamefont {Gustavsson}},\ and\ \bibinfo
  {author} {\bibfnamefont {W.~D.}\ \bibnamefont {Oliver}},\ }\bibinfo {title}
  {Superconducting Qubits: Current State of Play},\ \href
  {https://doi.org/https://doi.org/10.1146/annurev-conmatphys-031119-050605}
  {\bibfield  {journal} {\bibinfo  {journal} {Annu. Rev. Condens. Matter
  Phys.}\ }\textbf {\bibinfo {volume} {11}},\ \bibinfo {pages} {369} (\bibinfo
  {year} {2020})}\BibitemShut {NoStop}%
\bibitem [{\citenamefont {Siddiqi}(2021)}]{siddiqi2021engineering}%
  \BibitemOpen
  \bibfield  {author} {\bibinfo {author} {\bibfnamefont {I.}~\bibnamefont
  {Siddiqi}},\ }\bibinfo {title} {Engineering high-coherence superconducting
  qubits},\ \href {https://www.nature.com/articles/s41578-021-00370-4}
  {\bibfield  {journal} {\bibinfo  {journal} {Nat. Rev. Mater.}\ }\textbf
  {\bibinfo {volume} {6}},\ \bibinfo {pages} {875} (\bibinfo {year}
  {2021})}\BibitemShut {NoStop}%
\bibitem [{\citenamefont {Acharya}\ \emph {et~al.}(2024)\citenamefont
  {Acharya}, \citenamefont {Abanin}, \citenamefont {Aghababaie-Beni},
  \citenamefont {Aleiner}, \citenamefont {Andersen}, \citenamefont {Ansmann},
  \citenamefont {Arute}, \citenamefont {Arya}, \citenamefont {Asfaw},
  \citenamefont {Astrakhantsev}, \citenamefont {Atalaya}, \citenamefont
  {Babbush}, \citenamefont {Bacon}, \citenamefont {Ballard}, \citenamefont
  {Bardin}, \citenamefont {Bausch}, \citenamefont {Bengtsson}, \citenamefont
  {Bilmes}, \citenamefont {Blackwell}, \citenamefont {Boixo}, \citenamefont
  {Bortoli}, \citenamefont {Bourassa}, \citenamefont {Bovaird}, \citenamefont
  {Brill}, \citenamefont {Broughton}, \citenamefont {Browne}, \citenamefont
  {Buchea}, \citenamefont {Buckley}, \citenamefont {Buell}, \citenamefont
  {Burger}, \citenamefont {Burkett}, \citenamefont {Bushnell}, \citenamefont
  {Cabrera}, \citenamefont {Campero}, \citenamefont {Chang}, \citenamefont
  {Chen}, \citenamefont {Chen}, \citenamefont {Chiaro}, \citenamefont {Chik},
  \citenamefont {Chou}, \citenamefont {Claes}, \citenamefont {Cleland},
  \citenamefont {Cogan}, \citenamefont {Collins}, \citenamefont {Conner},
  \citenamefont {Courtney}, \citenamefont {Crook}, \citenamefont {Curtin},
  \citenamefont {Das}, \citenamefont {Davies}, \citenamefont {De~Lorenzo},
  \citenamefont {Debroy}, \citenamefont {Demura}, \citenamefont {Devoret},
  \citenamefont {Di~Paolo}, \citenamefont {Donohoe}, \citenamefont {Drozdov},
  \citenamefont {Dunsworth}, \citenamefont {Earle}, \citenamefont {Edlich},
  \citenamefont {Eickbusch}, \citenamefont {Elbag}, \citenamefont {Elzouka},
  \citenamefont {Erickson}, \citenamefont {Faoro}, \citenamefont {Farhi},
  \citenamefont {Ferreira}, \citenamefont {Burgos}, \citenamefont {Forati},
  \citenamefont {Fowler}, \citenamefont {Foxen}, \citenamefont {Ganjam},
  \citenamefont {Garcia}, \citenamefont {Gasca}, \citenamefont {Genois},
  \citenamefont {Giang}, \citenamefont {Gidney}, \citenamefont {Gilboa},
  \citenamefont {Gosula}, \citenamefont {Dau}, \citenamefont {Graumann},
  \citenamefont {Greene}, \citenamefont {Gross}, \citenamefont {Habegger},
  \citenamefont {Hall}, \citenamefont {Hamilton}, \citenamefont {Hansen},
  \citenamefont {Harrigan}, \citenamefont {Harrington}, \citenamefont {Heras},
  \citenamefont {Heslin}, \citenamefont {Heu}, \citenamefont {Higgott},
  \citenamefont {Hill}, \citenamefont {Hilton}, \citenamefont {Holland},
  \citenamefont {Hong}, \citenamefont {Huang}, \citenamefont {Huff},
  \citenamefont {Huggins}, \citenamefont {Ioffe}, \citenamefont {Isakov},
  \citenamefont {Iveland}, \citenamefont {Jeffrey}, \citenamefont {Jiang},
  \citenamefont {Jones}, \citenamefont {Jordan}, \citenamefont {Joshi},
  \citenamefont {Juhas}, \citenamefont {Kafri}, \citenamefont {Kang},
  \citenamefont {Karamlou}, \citenamefont {Kechedzhi}, \citenamefont {Kelly},
  \citenamefont {Khaire}, \citenamefont {Khattar}, \citenamefont {Khezri},
  \citenamefont {Kim}, \citenamefont {Klimov}, \citenamefont {Klots},
  \citenamefont {Kobrin}, \citenamefont {Kohli}, \citenamefont {Korotkov},
  \citenamefont {Kostritsa}, \citenamefont {Kothari}, \citenamefont
  {Kozlovskii}, \citenamefont {Kreikebaum}, \citenamefont {Kurilovich},
  \citenamefont {Lacroix}, \citenamefont {Landhuis}, \citenamefont {Lange-Dei},
  \citenamefont {Langley}, \citenamefont {Laptev}, \citenamefont {Lau},
  \citenamefont {Le~Guevel}, \citenamefont {Ledford}, \citenamefont {Lee},
  \citenamefont {Lee}, \citenamefont {Lensky}, \citenamefont {Leon},
  \citenamefont {Lester}, \citenamefont {Li}, \citenamefont {Li}, \citenamefont
  {Lill}, \citenamefont {Liu}, \citenamefont {Livingston}, \citenamefont
  {Locharla}, \citenamefont {Lucero}, \citenamefont {Lundahl}, \citenamefont
  {Lunt}, \citenamefont {Madhuk}, \citenamefont {Malone}, \citenamefont
  {Maloney}, \citenamefont {Mandra}, \citenamefont {Manyika}, \citenamefont
  {Martin}, \citenamefont {Martin}, \citenamefont {Martin}, \citenamefont
  {Maxfield}, \citenamefont {McClean}, \citenamefont {McEwen}, \citenamefont
  {Meeks}, \citenamefont {Megrant}, \citenamefont {Mi}, \citenamefont {Miao},
  \citenamefont {Mieszala}, \citenamefont {Molavi}, \citenamefont {Molina},
  \citenamefont {Montazeri}, \citenamefont {Morvan}, \citenamefont {Movassagh},
  \citenamefont {Mruczkiewicz}, \citenamefont {Naaman}, \citenamefont {Neeley},
  \citenamefont {Neill}, \citenamefont {Nersisyan}, \citenamefont {Neven},
  \citenamefont {Newman}, \citenamefont {Ng}, \citenamefont {Nguyen},
  \citenamefont {Nguyen}, \citenamefont {Ni}, \citenamefont {Niu},
  \citenamefont {O'Brien}, \citenamefont {Oliver}, \citenamefont {Opremcak},
  \citenamefont {Ottosson}, \citenamefont {Petukhov}, \citenamefont {Pizzuto},
  \citenamefont {Platt}, \citenamefont {Potter}, \citenamefont {Pritchard},
  \citenamefont {Pryadko}, \citenamefont {Quintana}, \citenamefont
  {Ramachandran}, \citenamefont {Reagor}, \citenamefont {Redding},
  \citenamefont {Rhodes}, \citenamefont {Roberts}, \citenamefont {Rosenberg},
  \citenamefont {Rosenfeld}, \citenamefont {Roushan}, \citenamefont {Rubin},
  \citenamefont {Saei}, \citenamefont {Sank}, \citenamefont {Sankaragomathi},
  \citenamefont {Satzinger}, \citenamefont {Schurkus}, \citenamefont
  {Schuster}, \citenamefont {Senior}, \citenamefont {Shearn}, \citenamefont
  {Shorter}, \citenamefont {Shutty}, \citenamefont {Shvarts}, \citenamefont
  {Singh}, \citenamefont {Sivak}, \citenamefont {Skruzny}, \citenamefont
  {Small}, \citenamefont {Smelyanskiy}, \citenamefont {Smith}, \citenamefont
  {Somma}, \citenamefont {Springer}, \citenamefont {Sterling}, \citenamefont
  {Strain}, \citenamefont {Suchard}, \citenamefont {Szasz}, \citenamefont
  {Sztein}, \citenamefont {Thor}, \citenamefont {Torres}, \citenamefont
  {Torunbalci}, \citenamefont {Vaishnav}, \citenamefont {Vargas}, \citenamefont
  {Vdovichev}, \citenamefont {Vidal}, \citenamefont {Villalonga}, \citenamefont
  {Heidweiller}, \citenamefont {Waltman}, \citenamefont {Wang}, \citenamefont
  {Ware}, \citenamefont {Weber}, \citenamefont {Weidel}, \citenamefont {White},
  \citenamefont {Wong}, \citenamefont {Woo}, \citenamefont {Xing},
  \citenamefont {Yao}, \citenamefont {Yeh}, \citenamefont {Ying}, \citenamefont
  {Yoo}, \citenamefont {Yosri}, \citenamefont {Young}, \citenamefont {Zalcman},
  \citenamefont {Zhang}, \citenamefont {Zhu}, \citenamefont {Zobrist},\ and\
  \citenamefont {Collaborato}}]{Google_EC_2025}%
  \BibitemOpen
  \bibfield  {author} {\bibinfo {author} {\bibfnamefont {R.}~\bibnamefont
  {Acharya}}, \bibinfo {author} {\bibfnamefont {D.~A.}\ \bibnamefont {Abanin}},
  \bibinfo {author} {\bibfnamefont {L.}~\bibnamefont {Aghababaie-Beni}},
  \bibinfo {author} {\bibfnamefont {I.}~\bibnamefont {Aleiner}}, \bibinfo
  {author} {\bibfnamefont {T.~I.}\ \bibnamefont {Andersen}}, \bibinfo {author}
  {\bibfnamefont {M.}~\bibnamefont {Ansmann}}, \bibinfo {author} {\bibfnamefont
  {F.}~\bibnamefont {Arute}}, \bibinfo {author} {\bibfnamefont
  {K.}~\bibnamefont {Arya}}, \bibinfo {author} {\bibfnamefont {A.}~\bibnamefont
  {Asfaw}}, \bibinfo {author} {\bibfnamefont {N.}~\bibnamefont
  {Astrakhantsev}}, \bibinfo {author} {\bibfnamefont {J.}~\bibnamefont
  {Atalaya}}, \bibinfo {author} {\bibfnamefont {R.}~\bibnamefont {Babbush}},
  \bibinfo {author} {\bibfnamefont {D.}~\bibnamefont {Bacon}}, \bibinfo
  {author} {\bibfnamefont {B.}~\bibnamefont {Ballard}}, \bibinfo {author}
  {\bibfnamefont {J.~C.}\ \bibnamefont {Bardin}}, \bibinfo {author}
  {\bibfnamefont {J.}~\bibnamefont {Bausch}}, \bibinfo {author} {\bibfnamefont
  {A.}~\bibnamefont {Bengtsson}}, \bibinfo {author} {\bibfnamefont
  {A.}~\bibnamefont {Bilmes}}, \bibinfo {author} {\bibfnamefont
  {S.}~\bibnamefont {Blackwell}}, \bibinfo {author} {\bibfnamefont
  {S.}~\bibnamefont {Boixo}}, \bibinfo {author} {\bibfnamefont
  {G.}~\bibnamefont {Bortoli}}, \bibinfo {author} {\bibfnamefont
  {A.}~\bibnamefont {Bourassa}}, \bibinfo {author} {\bibfnamefont
  {J.}~\bibnamefont {Bovaird}}, \bibinfo {author} {\bibfnamefont
  {L.}~\bibnamefont {Brill}}, \bibinfo {author} {\bibfnamefont
  {M.}~\bibnamefont {Broughton}}, \bibinfo {author} {\bibfnamefont {D.~A.}\
  \bibnamefont {Browne}}, \bibinfo {author} {\bibfnamefont {B.}~\bibnamefont
  {Buchea}}, \bibinfo {author} {\bibfnamefont {B.~B.}\ \bibnamefont {Buckley}},
  \bibinfo {author} {\bibfnamefont {D.~A.}\ \bibnamefont {Buell}}, \bibinfo
  {author} {\bibfnamefont {T.}~\bibnamefont {Burger}}, \bibinfo {author}
  {\bibfnamefont {B.}~\bibnamefont {Burkett}}, \bibinfo {author} {\bibfnamefont
  {N.}~\bibnamefont {Bushnell}}, \bibinfo {author} {\bibfnamefont
  {A.}~\bibnamefont {Cabrera}}, \bibinfo {author} {\bibfnamefont
  {J.}~\bibnamefont {Campero}}, \bibinfo {author} {\bibfnamefont {H.-S.}\
  \bibnamefont {Chang}}, \bibinfo {author} {\bibfnamefont {Y.}~\bibnamefont
  {Chen}}, \bibinfo {author} {\bibfnamefont {Z.}~\bibnamefont {Chen}}, \bibinfo
  {author} {\bibfnamefont {B.}~\bibnamefont {Chiaro}}, \bibinfo {author}
  {\bibfnamefont {D.}~\bibnamefont {Chik}}, \bibinfo {author} {\bibfnamefont
  {C.}~\bibnamefont {Chou}}, \bibinfo {author} {\bibfnamefont {J.}~\bibnamefont
  {Claes}}, \bibinfo {author} {\bibfnamefont {A.~Y.}\ \bibnamefont {Cleland}},
  \bibinfo {author} {\bibfnamefont {J.}~\bibnamefont {Cogan}}, \bibinfo
  {author} {\bibfnamefont {R.}~\bibnamefont {Collins}}, \bibinfo {author}
  {\bibfnamefont {P.}~\bibnamefont {Conner}}, \bibinfo {author} {\bibfnamefont
  {W.}~\bibnamefont {Courtney}}, \bibinfo {author} {\bibfnamefont {A.~L.}\
  \bibnamefont {Crook}}, \bibinfo {author} {\bibfnamefont {B.}~\bibnamefont
  {Curtin}}, \bibinfo {author} {\bibfnamefont {S.}~\bibnamefont {Das}},
  \bibinfo {author} {\bibfnamefont {A.}~\bibnamefont {Davies}}, \bibinfo
  {author} {\bibfnamefont {L.}~\bibnamefont {De~Lorenzo}}, \bibinfo {author}
  {\bibfnamefont {D.~M.}\ \bibnamefont {Debroy}}, \bibinfo {author}
  {\bibfnamefont {S.}~\bibnamefont {Demura}}, \bibinfo {author} {\bibfnamefont
  {M.}~\bibnamefont {Devoret}}, \bibinfo {author} {\bibfnamefont
  {A.}~\bibnamefont {Di~Paolo}}, \bibinfo {author} {\bibfnamefont
  {P.}~\bibnamefont {Donohoe}}, \bibinfo {author} {\bibfnamefont
  {I.}~\bibnamefont {Drozdov}}, \bibinfo {author} {\bibfnamefont
  {A.}~\bibnamefont {Dunsworth}}, \bibinfo {author} {\bibfnamefont
  {C.}~\bibnamefont {Earle}}, \bibinfo {author} {\bibfnamefont
  {T.}~\bibnamefont {Edlich}}, \bibinfo {author} {\bibfnamefont
  {A.}~\bibnamefont {Eickbusch}}, \bibinfo {author} {\bibfnamefont {A.~M.}\
  \bibnamefont {Elbag}}, \bibinfo {author} {\bibfnamefont {M.}~\bibnamefont
  {Elzouka}}, \bibinfo {author} {\bibfnamefont {C.}~\bibnamefont {Erickson}},
  \bibinfo {author} {\bibfnamefont {L.}~\bibnamefont {Faoro}}, \bibinfo
  {author} {\bibfnamefont {E.}~\bibnamefont {Farhi}}, \bibinfo {author}
  {\bibfnamefont {V.~S.}\ \bibnamefont {Ferreira}}, \bibinfo {author}
  {\bibfnamefont {L.~F.}\ \bibnamefont {Burgos}}, \bibinfo {author}
  {\bibfnamefont {E.}~\bibnamefont {Forati}}, \bibinfo {author} {\bibfnamefont
  {A.~G.}\ \bibnamefont {Fowler}}, \bibinfo {author} {\bibfnamefont
  {B.}~\bibnamefont {Foxen}}, \bibinfo {author} {\bibfnamefont
  {S.}~\bibnamefont {Ganjam}}, \bibinfo {author} {\bibfnamefont
  {G.}~\bibnamefont {Garcia}}, \bibinfo {author} {\bibfnamefont
  {R.}~\bibnamefont {Gasca}}, \bibinfo {author} {\bibfnamefont
  {E.}~\bibnamefont {Genois}}, \bibinfo {author} {\bibfnamefont
  {W.}~\bibnamefont {Giang}}, \bibinfo {author} {\bibfnamefont
  {C.}~\bibnamefont {Gidney}}, \bibinfo {author} {\bibfnamefont
  {D.}~\bibnamefont {Gilboa}}, \bibinfo {author} {\bibfnamefont
  {R.}~\bibnamefont {Gosula}}, \bibinfo {author} {\bibfnamefont {A.~G.}\
  \bibnamefont {Dau}}, \bibinfo {author} {\bibfnamefont {D.}~\bibnamefont
  {Graumann}}, \bibinfo {author} {\bibfnamefont {A.}~\bibnamefont {Greene}},
  \bibinfo {author} {\bibfnamefont {J.~A.}\ \bibnamefont {Gross}}, \bibinfo
  {author} {\bibfnamefont {S.}~\bibnamefont {Habegger}}, \bibinfo {author}
  {\bibfnamefont {J.}~\bibnamefont {Hall}}, \bibinfo {author} {\bibfnamefont
  {M.~C.}\ \bibnamefont {Hamilton}}, \bibinfo {author} {\bibfnamefont
  {M.}~\bibnamefont {Hansen}}, \bibinfo {author} {\bibfnamefont {M.~P.}\
  \bibnamefont {Harrigan}}, \bibinfo {author} {\bibfnamefont {S.~D.}\
  \bibnamefont {Harrington}}, \bibinfo {author} {\bibfnamefont {F.~J.~H.}\
  \bibnamefont {Heras}}, \bibinfo {author} {\bibfnamefont {S.}~\bibnamefont
  {Heslin}}, \bibinfo {author} {\bibfnamefont {P.}~\bibnamefont {Heu}},
  \bibinfo {author} {\bibfnamefont {O.}~\bibnamefont {Higgott}}, \bibinfo
  {author} {\bibfnamefont {G.}~\bibnamefont {Hill}}, \bibinfo {author}
  {\bibfnamefont {J.}~\bibnamefont {Hilton}}, \bibinfo {author} {\bibfnamefont
  {G.}~\bibnamefont {Holland}}, \bibinfo {author} {\bibfnamefont
  {S.}~\bibnamefont {Hong}}, \bibinfo {author} {\bibfnamefont {H.-Y.}\
  \bibnamefont {Huang}}, \bibinfo {author} {\bibfnamefont {A.}~\bibnamefont
  {Huff}}, \bibinfo {author} {\bibfnamefont {W.~J.}\ \bibnamefont {Huggins}},
  \bibinfo {author} {\bibfnamefont {L.~B.}\ \bibnamefont {Ioffe}}, \bibinfo
  {author} {\bibfnamefont {S.~V.}\ \bibnamefont {Isakov}}, \bibinfo {author}
  {\bibfnamefont {J.}~\bibnamefont {Iveland}}, \bibinfo {author} {\bibfnamefont
  {E.}~\bibnamefont {Jeffrey}}, \bibinfo {author} {\bibfnamefont
  {Z.}~\bibnamefont {Jiang}}, \bibinfo {author} {\bibfnamefont
  {C.}~\bibnamefont {Jones}}, \bibinfo {author} {\bibfnamefont
  {S.}~\bibnamefont {Jordan}}, \bibinfo {author} {\bibfnamefont
  {C.}~\bibnamefont {Joshi}}, \bibinfo {author} {\bibfnamefont
  {P.}~\bibnamefont {Juhas}}, \bibinfo {author} {\bibfnamefont
  {D.}~\bibnamefont {Kafri}}, \bibinfo {author} {\bibfnamefont
  {H.}~\bibnamefont {Kang}}, \bibinfo {author} {\bibfnamefont {A.~H.}\
  \bibnamefont {Karamlou}}, \bibinfo {author} {\bibfnamefont {K.}~\bibnamefont
  {Kechedzhi}}, \bibinfo {author} {\bibfnamefont {J.}~\bibnamefont {Kelly}},
  \bibinfo {author} {\bibfnamefont {T.}~\bibnamefont {Khaire}}, \bibinfo
  {author} {\bibfnamefont {T.}~\bibnamefont {Khattar}}, \bibinfo {author}
  {\bibfnamefont {M.}~\bibnamefont {Khezri}}, \bibinfo {author} {\bibfnamefont
  {S.}~\bibnamefont {Kim}}, \bibinfo {author} {\bibfnamefont {P.~V.}\
  \bibnamefont {Klimov}}, \bibinfo {author} {\bibfnamefont {A.~R.}\
  \bibnamefont {Klots}}, \bibinfo {author} {\bibfnamefont {B.}~\bibnamefont
  {Kobrin}}, \bibinfo {author} {\bibfnamefont {P.}~\bibnamefont {Kohli}},
  \bibinfo {author} {\bibfnamefont {A.~N.}\ \bibnamefont {Korotkov}}, \bibinfo
  {author} {\bibfnamefont {F.}~\bibnamefont {Kostritsa}}, \bibinfo {author}
  {\bibfnamefont {R.}~\bibnamefont {Kothari}}, \bibinfo {author} {\bibfnamefont
  {B.}~\bibnamefont {Kozlovskii}}, \bibinfo {author} {\bibfnamefont {J.~M.}\
  \bibnamefont {Kreikebaum}}, \bibinfo {author} {\bibfnamefont {V.~D.}\
  \bibnamefont {Kurilovich}}, \bibinfo {author} {\bibfnamefont
  {N.}~\bibnamefont {Lacroix}}, \bibinfo {author} {\bibfnamefont
  {D.}~\bibnamefont {Landhuis}}, \bibinfo {author} {\bibfnamefont
  {T.}~\bibnamefont {Lange-Dei}}, \bibinfo {author} {\bibfnamefont {B.~W.}\
  \bibnamefont {Langley}}, \bibinfo {author} {\bibfnamefont {P.}~\bibnamefont
  {Laptev}}, \bibinfo {author} {\bibfnamefont {K.-M.}\ \bibnamefont {Lau}},
  \bibinfo {author} {\bibfnamefont {L.}~\bibnamefont {Le~Guevel}}, \bibinfo
  {author} {\bibfnamefont {J.}~\bibnamefont {Ledford}}, \bibinfo {author}
  {\bibfnamefont {J.}~\bibnamefont {Lee}}, \bibinfo {author} {\bibfnamefont
  {K.}~\bibnamefont {Lee}}, \bibinfo {author} {\bibfnamefont {Y.~D.}\
  \bibnamefont {Lensky}}, \bibinfo {author} {\bibfnamefont {S.}~\bibnamefont
  {Leon}}, \bibinfo {author} {\bibfnamefont {B.~J.}\ \bibnamefont {Lester}},
  \bibinfo {author} {\bibfnamefont {W.~Y.}\ \bibnamefont {Li}}, \bibinfo
  {author} {\bibfnamefont {Y.}~\bibnamefont {Li}}, \bibinfo {author}
  {\bibfnamefont {A.~T.}\ \bibnamefont {Lill}}, \bibinfo {author}
  {\bibfnamefont {W.}~\bibnamefont {Liu}}, \bibinfo {author} {\bibfnamefont
  {W.~P.}\ \bibnamefont {Livingston}}, \bibinfo {author} {\bibfnamefont
  {A.}~\bibnamefont {Locharla}}, \bibinfo {author} {\bibfnamefont
  {E.}~\bibnamefont {Lucero}}, \bibinfo {author} {\bibfnamefont
  {D.}~\bibnamefont {Lundahl}}, \bibinfo {author} {\bibfnamefont
  {A.}~\bibnamefont {Lunt}}, \bibinfo {author} {\bibfnamefont {S.}~\bibnamefont
  {Madhuk}}, \bibinfo {author} {\bibfnamefont {F.~D.}\ \bibnamefont {Malone}},
  \bibinfo {author} {\bibfnamefont {A.}~\bibnamefont {Maloney}}, \bibinfo
  {author} {\bibfnamefont {S.}~\bibnamefont {Mandra}}, \bibinfo {author}
  {\bibfnamefont {J.}~\bibnamefont {Manyika}}, \bibinfo {author} {\bibfnamefont
  {L.~S.}\ \bibnamefont {Martin}}, \bibinfo {author} {\bibfnamefont
  {O.}~\bibnamefont {Martin}}, \bibinfo {author} {\bibfnamefont
  {S.}~\bibnamefont {Martin}}, \bibinfo {author} {\bibfnamefont
  {C.}~\bibnamefont {Maxfield}}, \bibinfo {author} {\bibfnamefont {J.~R.}\
  \bibnamefont {McClean}}, \bibinfo {author} {\bibfnamefont {M.}~\bibnamefont
  {McEwen}}, \bibinfo {author} {\bibfnamefont {S.}~\bibnamefont {Meeks}},
  \bibinfo {author} {\bibfnamefont {A.}~\bibnamefont {Megrant}}, \bibinfo
  {author} {\bibfnamefont {X.}~\bibnamefont {Mi}}, \bibinfo {author}
  {\bibfnamefont {K.~C.}\ \bibnamefont {Miao}}, \bibinfo {author}
  {\bibfnamefont {A.}~\bibnamefont {Mieszala}}, \bibinfo {author}
  {\bibfnamefont {R.}~\bibnamefont {Molavi}}, \bibinfo {author} {\bibfnamefont
  {S.}~\bibnamefont {Molina}}, \bibinfo {author} {\bibfnamefont
  {S.}~\bibnamefont {Montazeri}}, \bibinfo {author} {\bibfnamefont
  {A.}~\bibnamefont {Morvan}}, \bibinfo {author} {\bibfnamefont
  {R.}~\bibnamefont {Movassagh}}, \bibinfo {author} {\bibfnamefont
  {W.}~\bibnamefont {Mruczkiewicz}}, \bibinfo {author} {\bibfnamefont
  {O.}~\bibnamefont {Naaman}}, \bibinfo {author} {\bibfnamefont
  {M.}~\bibnamefont {Neeley}}, \bibinfo {author} {\bibfnamefont
  {C.}~\bibnamefont {Neill}}, \bibinfo {author} {\bibfnamefont
  {A.}~\bibnamefont {Nersisyan}}, \bibinfo {author} {\bibfnamefont
  {H.}~\bibnamefont {Neven}}, \bibinfo {author} {\bibfnamefont
  {M.}~\bibnamefont {Newman}}, \bibinfo {author} {\bibfnamefont {J.~H.}\
  \bibnamefont {Ng}}, \bibinfo {author} {\bibfnamefont {A.}~\bibnamefont
  {Nguyen}}, \bibinfo {author} {\bibfnamefont {M.}~\bibnamefont {Nguyen}},
  \bibinfo {author} {\bibfnamefont {C.-H.}\ \bibnamefont {Ni}}, \bibinfo
  {author} {\bibfnamefont {M.~Y.}\ \bibnamefont {Niu}}, \bibinfo {author}
  {\bibfnamefont {T.~E.}\ \bibnamefont {O'Brien}}, \bibinfo {author}
  {\bibfnamefont {W.~D.}\ \bibnamefont {Oliver}}, \bibinfo {author}
  {\bibfnamefont {A.}~\bibnamefont {Opremcak}}, \bibinfo {author}
  {\bibfnamefont {K.}~\bibnamefont {Ottosson}}, \bibinfo {author}
  {\bibfnamefont {A.}~\bibnamefont {Petukhov}}, \bibinfo {author}
  {\bibfnamefont {A.}~\bibnamefont {Pizzuto}}, \bibinfo {author} {\bibfnamefont
  {J.}~\bibnamefont {Platt}}, \bibinfo {author} {\bibfnamefont
  {R.}~\bibnamefont {Potter}}, \bibinfo {author} {\bibfnamefont
  {O.}~\bibnamefont {Pritchard}}, \bibinfo {author} {\bibfnamefont {L.~P.}\
  \bibnamefont {Pryadko}}, \bibinfo {author} {\bibfnamefont {C.}~\bibnamefont
  {Quintana}}, \bibinfo {author} {\bibfnamefont {G.}~\bibnamefont
  {Ramachandran}}, \bibinfo {author} {\bibfnamefont {M.~J.}\ \bibnamefont
  {Reagor}}, \bibinfo {author} {\bibfnamefont {J.}~\bibnamefont {Redding}},
  \bibinfo {author} {\bibfnamefont {D.~M.}\ \bibnamefont {Rhodes}}, \bibinfo
  {author} {\bibfnamefont {G.}~\bibnamefont {Roberts}}, \bibinfo {author}
  {\bibfnamefont {E.}~\bibnamefont {Rosenberg}}, \bibinfo {author}
  {\bibfnamefont {E.}~\bibnamefont {Rosenfeld}}, \bibinfo {author}
  {\bibfnamefont {P.}~\bibnamefont {Roushan}}, \bibinfo {author} {\bibfnamefont
  {N.~C.}\ \bibnamefont {Rubin}}, \bibinfo {author} {\bibfnamefont
  {N.}~\bibnamefont {Saei}}, \bibinfo {author} {\bibfnamefont {D.}~\bibnamefont
  {Sank}}, \bibinfo {author} {\bibfnamefont {K.}~\bibnamefont
  {Sankaragomathi}}, \bibinfo {author} {\bibfnamefont {K.~J.}\ \bibnamefont
  {Satzinger}}, \bibinfo {author} {\bibfnamefont {H.~F.}\ \bibnamefont
  {Schurkus}}, \bibinfo {author} {\bibfnamefont {C.}~\bibnamefont {Schuster}},
  \bibinfo {author} {\bibfnamefont {A.~W.}\ \bibnamefont {Senior}}, \bibinfo
  {author} {\bibfnamefont {M.~J.}\ \bibnamefont {Shearn}}, \bibinfo {author}
  {\bibfnamefont {A.}~\bibnamefont {Shorter}}, \bibinfo {author} {\bibfnamefont
  {N.}~\bibnamefont {Shutty}}, \bibinfo {author} {\bibfnamefont
  {V.}~\bibnamefont {Shvarts}}, \bibinfo {author} {\bibfnamefont
  {S.}~\bibnamefont {Singh}}, \bibinfo {author} {\bibfnamefont
  {V.}~\bibnamefont {Sivak}}, \bibinfo {author} {\bibfnamefont
  {J.}~\bibnamefont {Skruzny}}, \bibinfo {author} {\bibfnamefont
  {S.}~\bibnamefont {Small}}, \bibinfo {author} {\bibfnamefont
  {V.}~\bibnamefont {Smelyanskiy}}, \bibinfo {author} {\bibfnamefont {W.~C.}\
  \bibnamefont {Smith}}, \bibinfo {author} {\bibfnamefont {R.~D.}\ \bibnamefont
  {Somma}}, \bibinfo {author} {\bibfnamefont {S.}~\bibnamefont {Springer}},
  \bibinfo {author} {\bibfnamefont {G.}~\bibnamefont {Sterling}}, \bibinfo
  {author} {\bibfnamefont {D.}~\bibnamefont {Strain}}, \bibinfo {author}
  {\bibfnamefont {J.}~\bibnamefont {Suchard}}, \bibinfo {author} {\bibfnamefont
  {A.}~\bibnamefont {Szasz}}, \bibinfo {author} {\bibfnamefont
  {A.}~\bibnamefont {Sztein}}, \bibinfo {author} {\bibfnamefont
  {D.}~\bibnamefont {Thor}}, \bibinfo {author} {\bibfnamefont {A.}~\bibnamefont
  {Torres}}, \bibinfo {author} {\bibfnamefont {M.~M.}\ \bibnamefont
  {Torunbalci}}, \bibinfo {author} {\bibfnamefont {A.}~\bibnamefont
  {Vaishnav}}, \bibinfo {author} {\bibfnamefont {J.}~\bibnamefont {Vargas}},
  \bibinfo {author} {\bibfnamefont {S.}~\bibnamefont {Vdovichev}}, \bibinfo
  {author} {\bibfnamefont {G.}~\bibnamefont {Vidal}}, \bibinfo {author}
  {\bibfnamefont {B.}~\bibnamefont {Villalonga}}, \bibinfo {author}
  {\bibfnamefont {C.~V.}\ \bibnamefont {Heidweiller}}, \bibinfo {author}
  {\bibfnamefont {S.}~\bibnamefont {Waltman}}, \bibinfo {author} {\bibfnamefont
  {S.~X.}\ \bibnamefont {Wang}}, \bibinfo {author} {\bibfnamefont
  {B.}~\bibnamefont {Ware}}, \bibinfo {author} {\bibfnamefont {K.}~\bibnamefont
  {Weber}}, \bibinfo {author} {\bibfnamefont {T.}~\bibnamefont {Weidel}},
  \bibinfo {author} {\bibfnamefont {T.}~\bibnamefont {White}}, \bibinfo
  {author} {\bibfnamefont {K.}~\bibnamefont {Wong}}, \bibinfo {author}
  {\bibfnamefont {B.~W.~K.}\ \bibnamefont {Woo}}, \bibinfo {author}
  {\bibfnamefont {C.}~\bibnamefont {Xing}}, \bibinfo {author} {\bibfnamefont
  {Z.~J.}\ \bibnamefont {Yao}}, \bibinfo {author} {\bibfnamefont
  {P.}~\bibnamefont {Yeh}}, \bibinfo {author} {\bibfnamefont {B.}~\bibnamefont
  {Ying}}, \bibinfo {author} {\bibfnamefont {J.}~\bibnamefont {Yoo}}, \bibinfo
  {author} {\bibfnamefont {N.}~\bibnamefont {Yosri}}, \bibinfo {author}
  {\bibfnamefont {G.}~\bibnamefont {Young}}, \bibinfo {author} {\bibfnamefont
  {A.}~\bibnamefont {Zalcman}}, \bibinfo {author} {\bibfnamefont
  {Y.}~\bibnamefont {Zhang}}, \bibinfo {author} {\bibfnamefont
  {N.}~\bibnamefont {Zhu}}, \bibinfo {author} {\bibfnamefont {N.}~\bibnamefont
  {Zobrist}},\ and\ \bibinfo {author} {\bibfnamefont {G.~Q. A.~.}\ \bibnamefont
  {Collaborato}},\ }\bibinfo {title} {Quantum error correction below the
  surface code threshold},\ \href {https://doi.org/10.1038/s41586-024-08449-y}
  {\bibfield  {journal} {\bibinfo  {journal} {Nature}\ }\textbf {\bibinfo
  {volume} {638}},\ \bibinfo {pages} {920} (\bibinfo {year}
  {2024})}\BibitemShut {NoStop}%
\bibitem [{\citenamefont {Li}\ \emph {et~al.}(2024)\citenamefont {Li},
  \citenamefont {Xu}, \citenamefont {Wang}, \citenamefont {Tang}, \citenamefont
  {Zhang}, \citenamefont {Yang}, \citenamefont {Su}, \citenamefont {Wang},
  \citenamefont {Mi}, \citenamefont {Sun}, \citenamefont {Liang}, \citenamefont
  {Chen}, \citenamefont {Li}, \citenamefont {Zhang}, \citenamefont {Linghu},
  \citenamefont {Han}, \citenamefont {Liu}, \citenamefont {Feng}, \citenamefont
  {Liu}, \citenamefont {Xue}, \citenamefont {Zhang}, \citenamefont {Jin},
  \citenamefont {Zhu}, \citenamefont {Yu}, \citenamefont {Zhao},\ and\
  \citenamefont {Xue}}]{fab}%
  \BibitemOpen
  \bibfield  {author} {\bibinfo {author} {\bibfnamefont {X.}~\bibnamefont
  {Li}}, \bibinfo {author} {\bibfnamefont {H.}~\bibnamefont {Xu}}, \bibinfo
  {author} {\bibfnamefont {J.}~\bibnamefont {Wang}}, \bibinfo {author}
  {\bibfnamefont {L.-Z.}\ \bibnamefont {Tang}}, \bibinfo {author}
  {\bibfnamefont {D.-W.}\ \bibnamefont {Zhang}}, \bibinfo {author}
  {\bibfnamefont {C.}~\bibnamefont {Yang}}, \bibinfo {author} {\bibfnamefont
  {T.}~\bibnamefont {Su}}, \bibinfo {author} {\bibfnamefont {C.}~\bibnamefont
  {Wang}}, \bibinfo {author} {\bibfnamefont {Z.}~\bibnamefont {Mi}}, \bibinfo
  {author} {\bibfnamefont {W.}~\bibnamefont {Sun}}, \bibinfo {author}
  {\bibfnamefont {X.}~\bibnamefont {Liang}}, \bibinfo {author} {\bibfnamefont
  {M.}~\bibnamefont {Chen}}, \bibinfo {author} {\bibfnamefont {C.}~\bibnamefont
  {Li}}, \bibinfo {author} {\bibfnamefont {Y.}~\bibnamefont {Zhang}}, \bibinfo
  {author} {\bibfnamefont {K.}~\bibnamefont {Linghu}}, \bibinfo {author}
  {\bibfnamefont {J.}~\bibnamefont {Han}}, \bibinfo {author} {\bibfnamefont
  {W.}~\bibnamefont {Liu}}, \bibinfo {author} {\bibfnamefont {Y.}~\bibnamefont
  {Feng}}, \bibinfo {author} {\bibfnamefont {P.}~\bibnamefont {Liu}}, \bibinfo
  {author} {\bibfnamefont {G.}~\bibnamefont {Xue}}, \bibinfo {author}
  {\bibfnamefont {J.}~\bibnamefont {Zhang}}, \bibinfo {author} {\bibfnamefont
  {Y.}~\bibnamefont {Jin}}, \bibinfo {author} {\bibfnamefont {S.-L.}\
  \bibnamefont {Zhu}}, \bibinfo {author} {\bibfnamefont {H.}~\bibnamefont
  {Yu}}, \bibinfo {author} {\bibfnamefont {S.~P.}\ \bibnamefont {Zhao}},\ and\
  \bibinfo {author} {\bibfnamefont {Q.-K.}\ \bibnamefont {Xue}},\ }\bibinfo
  {title} {Mapping the topology-localization phase diagram with quasiperiodic
  disorder using a programmable superconducting simulator},\ \href
  {https://doi.org/10.1103/PhysRevResearch.6.L042038} {\bibfield  {journal}
  {\bibinfo  {journal} {Phys. Rev. Res.}\ }\textbf {\bibinfo {volume} {6}},\
  \bibinfo {pages} {L042038} (\bibinfo {year} {2024})}\BibitemShut {NoStop}%
\bibitem [{\citenamefont {Li}\ \emph {et~al.}(2021)\citenamefont {Li},
  \citenamefont {Zhang}, \citenamefont {Yang}, \citenamefont {Li},
  \citenamefont {Wang}, \citenamefont {Su}, \citenamefont {Chen}, \citenamefont
  {Li}, \citenamefont {Li}, \citenamefont {Mi}, \citenamefont {Liang},
  \citenamefont {Wang}, \citenamefont {Yang}, \citenamefont {Feng},
  \citenamefont {Linghu}, \citenamefont {Xu}, \citenamefont {Han},
  \citenamefont {Liu}, \citenamefont {Zhao}, \citenamefont {Ma}, \citenamefont
  {Wang}, \citenamefont {Zhang}, \citenamefont {Song}, \citenamefont {Liu},
  \citenamefont {Wang}, \citenamefont {Yang}, \citenamefont {Xue},
  \citenamefont {Jin},\ and\ \citenamefont {Yu}}]{fab2}%
  \BibitemOpen
  \bibfield  {author} {\bibinfo {author} {\bibfnamefont {X.}~\bibnamefont
  {Li}}, \bibinfo {author} {\bibfnamefont {Y.}~\bibnamefont {Zhang}}, \bibinfo
  {author} {\bibfnamefont {C.}~\bibnamefont {Yang}}, \bibinfo {author}
  {\bibfnamefont {Z.}~\bibnamefont {Li}}, \bibinfo {author} {\bibfnamefont
  {J.}~\bibnamefont {Wang}}, \bibinfo {author} {\bibfnamefont {T.}~\bibnamefont
  {Su}}, \bibinfo {author} {\bibfnamefont {M.}~\bibnamefont {Chen}}, \bibinfo
  {author} {\bibfnamefont {Y.}~\bibnamefont {Li}}, \bibinfo {author}
  {\bibfnamefont {C.}~\bibnamefont {Li}}, \bibinfo {author} {\bibfnamefont
  {Z.}~\bibnamefont {Mi}}, \bibinfo {author} {\bibfnamefont {X.}~\bibnamefont
  {Liang}}, \bibinfo {author} {\bibfnamefont {C.}~\bibnamefont {Wang}},
  \bibinfo {author} {\bibfnamefont {Z.}~\bibnamefont {Yang}}, \bibinfo {author}
  {\bibfnamefont {Y.}~\bibnamefont {Feng}}, \bibinfo {author} {\bibfnamefont
  {K.}~\bibnamefont {Linghu}}, \bibinfo {author} {\bibfnamefont
  {H.}~\bibnamefont {Xu}}, \bibinfo {author} {\bibfnamefont {J.}~\bibnamefont
  {Han}}, \bibinfo {author} {\bibfnamefont {W.}~\bibnamefont {Liu}}, \bibinfo
  {author} {\bibfnamefont {P.}~\bibnamefont {Zhao}}, \bibinfo {author}
  {\bibfnamefont {T.}~\bibnamefont {Ma}}, \bibinfo {author} {\bibfnamefont
  {R.}~\bibnamefont {Wang}}, \bibinfo {author} {\bibfnamefont {J.}~\bibnamefont
  {Zhang}}, \bibinfo {author} {\bibfnamefont {Y.}~\bibnamefont {Song}},
  \bibinfo {author} {\bibfnamefont {P.}~\bibnamefont {Liu}}, \bibinfo {author}
  {\bibfnamefont {Z.}~\bibnamefont {Wang}}, \bibinfo {author} {\bibfnamefont
  {Z.}~\bibnamefont {Yang}}, \bibinfo {author} {\bibfnamefont {G.}~\bibnamefont
  {Xue}}, \bibinfo {author} {\bibfnamefont {Y.}~\bibnamefont {Jin}},\ and\
  \bibinfo {author} {\bibfnamefont {H.}~\bibnamefont {Yu}},\ }\bibinfo {title}
  {Vacuum-gap transmon qubits realized using flip-chip technology},\ \href
  {https://doi.org/10.1063/5.0068255} {\bibfield  {journal} {\bibinfo
  {journal} {Appl. Phys. Lett.}\ }\textbf {\bibinfo {volume} {119}},\ \bibinfo
  {pages} {184003} (\bibinfo {year} {2021})}\BibitemShut {NoStop}%
\bibitem [{\citenamefont {Wang}\ \emph
  {et~al.}(2024{\natexlab{a}})\citenamefont {Wang}, \citenamefont {Chen},
  \citenamefont {Du}, \citenamefont {Yang}, \citenamefont {Cai}, \citenamefont
  {Huang}, \citenamefont {Zhang}, \citenamefont {Xu}, \citenamefont {Du},
  \citenamefont {Li} \emph {et~al.}}]{wang_quantum_2024}%
  \BibitemOpen
  \bibfield  {author} {\bibinfo {author} {\bibfnamefont {Z.}~\bibnamefont
  {Wang}}, \bibinfo {author} {\bibfnamefont {Q.}~\bibnamefont {Chen}}, \bibinfo
  {author} {\bibfnamefont {Y.}~\bibnamefont {Du}}, \bibinfo {author}
  {\bibfnamefont {Z.}~\bibnamefont {Yang}}, \bibinfo {author} {\bibfnamefont
  {X.}~\bibnamefont {Cai}}, \bibinfo {author} {\bibfnamefont {K.}~\bibnamefont
  {Huang}}, \bibinfo {author} {\bibfnamefont {J.}~\bibnamefont {Zhang}},
  \bibinfo {author} {\bibfnamefont {K.}~\bibnamefont {Xu}}, \bibinfo {author}
  {\bibfnamefont {J.}~\bibnamefont {Du}}, \bibinfo {author} {\bibfnamefont
  {Y.}~\bibnamefont {Li}}, \emph {et~al.},\ }\bibinfo {title} {Quantum
  compiling with reinforcement learning on a superconducting processor},\ \href
  {https://arxiv.org/abs/2406.12195} {\bibfield  {journal} {\bibinfo  {journal}
  {arXiv:2406.12195}\ } (\bibinfo {year} {2024}{\natexlab{a}})}\BibitemShut
  {NoStop}%
\bibitem [{\citenamefont {Wang}\ \emph
  {et~al.}(2024{\natexlab{b}})\citenamefont {Wang}, \citenamefont {Wang},
  \citenamefont {Zhao}, \citenamefont {Yang}, \citenamefont {Shi},
  \citenamefont {Huang}, \citenamefont {Xu}, \citenamefont {Zhang},
  \citenamefont {Fan}, \citenamefont {Zhao}, \citenamefont {Hu},\ and\
  \citenamefont {Yu}}]{wang_demonstration_2024}%
  \BibitemOpen
  \bibfield  {author} {\bibinfo {author} {\bibfnamefont {Z.~T.}\ \bibnamefont
  {Wang}}, \bibinfo {author} {\bibfnamefont {R.}~\bibnamefont {Wang}}, \bibinfo
  {author} {\bibfnamefont {P.}~\bibnamefont {Zhao}}, \bibinfo {author}
  {\bibfnamefont {Z.~H.}\ \bibnamefont {Yang}}, \bibinfo {author}
  {\bibfnamefont {Y.-H.}\ \bibnamefont {Shi}}, \bibinfo {author} {\bibfnamefont
  {K.}~\bibnamefont {Huang}}, \bibinfo {author} {\bibfnamefont
  {K.}~\bibnamefont {Xu}}, \bibinfo {author} {\bibfnamefont {Y.-S.}\
  \bibnamefont {Zhang}}, \bibinfo {author} {\bibfnamefont {H.}~\bibnamefont
  {Fan}}, \bibinfo {author} {\bibfnamefont {S.~P.}\ \bibnamefont {Zhao}},
  \bibinfo {author} {\bibfnamefont {M.-J.}\ \bibnamefont {Hu}},\ and\ \bibinfo
  {author} {\bibfnamefont {H.}~\bibnamefont {Yu}},\ }\bibinfo {title}
  {Demonstration of Maxwell demon-assisted Einstein-Podolsky-Rosen steering via
  superconducting quantum processor},\ \href
  {https://doi.org/10.1103/PhysRevResearch.6.L032073} {\bibfield  {journal}
  {\bibinfo  {journal} {Phys. Rev. Res.}\ }\textbf {\bibinfo {volume} {6}},\
  \bibinfo {pages} {L032073} (\bibinfo {year}
  {2024}{\natexlab{b}})}\BibitemShut {NoStop}%
\bibitem [{\citenamefont {Borla}\ \emph {et~al.}(2021)\citenamefont {Borla},
  \citenamefont {Verresen}, \citenamefont {Shah},\ and\ \citenamefont
  {Moroz}}]{10.21468/SciPostPhys.10.6.148}%
  \BibitemOpen
  \bibfield  {author} {\bibinfo {author} {\bibfnamefont {U.}~\bibnamefont
  {Borla}}, \bibinfo {author} {\bibfnamefont {R.}~\bibnamefont {Verresen}},
  \bibinfo {author} {\bibfnamefont {J.}~\bibnamefont {Shah}},\ and\ \bibinfo
  {author} {\bibfnamefont {S.}~\bibnamefont {Moroz}},\ }\bibinfo {title}
  {Gauging the Kitaev chain},\ \href
  {https://doi.org/10.21468/SciPostPhys.10.6.148} {\bibfield  {journal}
  {\bibinfo  {journal} {SciPost Phys.}\ }\textbf {\bibinfo {volume} {10}},\
  \bibinfo {pages} {148} (\bibinfo {year} {2021})}\BibitemShut {NoStop}%
\bibitem [{\citenamefont {Khemani}\ \emph {et~al.}(2016)\citenamefont
  {Khemani}, \citenamefont {Lazarides}, \citenamefont {Moessner},\ and\
  \citenamefont {Sondhi}}]{PhysRevLett.116.250401}%
  \BibitemOpen
  \bibfield  {author} {\bibinfo {author} {\bibfnamefont {V.}~\bibnamefont
  {Khemani}}, \bibinfo {author} {\bibfnamefont {A.}~\bibnamefont {Lazarides}},
  \bibinfo {author} {\bibfnamefont {R.}~\bibnamefont {Moessner}},\ and\
  \bibinfo {author} {\bibfnamefont {S.~L.}\ \bibnamefont {Sondhi}},\ }\bibinfo
  {title} {Phase Structure of Driven Quantum Systems},\ \href
  {https://doi.org/10.1103/PhysRevLett.116.250401} {\bibfield  {journal}
  {\bibinfo  {journal} {Phys. Rev. Lett.}\ }\textbf {\bibinfo {volume} {116}},\
  \bibinfo {pages} {250401} (\bibinfo {year} {2016})}\BibitemShut {NoStop}%
\bibitem [{\citenamefont {von Keyserlingk}\ and\ \citenamefont
  {Sondhi}(2016)}]{PhysRevB.93.245146}%
  \BibitemOpen
  \bibfield  {author} {\bibinfo {author} {\bibfnamefont {C.~W.}\ \bibnamefont
  {von Keyserlingk}}\ and\ \bibinfo {author} {\bibfnamefont {S.~L.}\
  \bibnamefont {Sondhi}},\ }\bibinfo {title} {Phase structure of
  one-dimensional interacting Floquet systems. II. Symmetry-broken phases},\
  \href {https://doi.org/10.1103/PhysRevB.93.245146} {\bibfield  {journal}
  {\bibinfo  {journal} {Phys. Rev. B}\ }\textbf {\bibinfo {volume} {93}},\
  \bibinfo {pages} {245146} (\bibinfo {year} {2016})}\BibitemShut {NoStop}%
\bibitem [{\citenamefont {Harper}\ \emph {et~al.}(2020)\citenamefont {Harper},
  \citenamefont {Roy}, \citenamefont {Rudner},\ and\ \citenamefont
  {Sondhi}}]{annurev:/content/journals/10.1146/annurev-conmatphys-031218-013721}%
  \BibitemOpen
  \bibfield  {author} {\bibinfo {author} {\bibfnamefont {F.}~\bibnamefont
  {Harper}}, \bibinfo {author} {\bibfnamefont {R.}~\bibnamefont {Roy}},
  \bibinfo {author} {\bibfnamefont {M.~S.}\ \bibnamefont {Rudner}},\ and\
  \bibinfo {author} {\bibfnamefont {S.}~\bibnamefont {Sondhi}},\ }\bibinfo
  {title} {Topology and Broken Symmetry in Floquet Systems},\ \href
  {https://doi.org/https://doi.org/10.1146/annurev-conmatphys-031218-013721}
  {\bibfield  {journal} {\bibinfo  {journal} {Annu. Rev. Condens. Matter
  Phys.}\ }\textbf {\bibinfo {volume} {11}},\ \bibinfo {pages} {345} (\bibinfo
  {year} {2020})}\BibitemShut {NoStop}%
\bibitem [{\citenamefont {Wannier}(1962)}]{RevModPhys.34.645}%
  \BibitemOpen
  \bibfield  {author} {\bibinfo {author} {\bibfnamefont {G.~H.}\ \bibnamefont
  {Wannier}},\ }\bibinfo {title} {Dynamics of Band Electrons in Electric and
  Magnetic Fields},\ \href {https://doi.org/10.1103/RevModPhys.34.645}
  {\bibfield  {journal} {\bibinfo  {journal} {Rev. Mod. Phys.}\ }\textbf
  {\bibinfo {volume} {34}},\ \bibinfo {pages} {645} (\bibinfo {year}
  {1962})}\BibitemShut {NoStop}%
\bibitem [{\citenamefont {Feldmann}\ \emph {et~al.}(1992)\citenamefont
  {Feldmann}, \citenamefont {Leo}, \citenamefont {Shah}, \citenamefont
  {Miller}, \citenamefont {Cunningham}, \citenamefont {Meier}, \citenamefont
  {von Plessen}, \citenamefont {Schulze}, \citenamefont {Thomas},\ and\
  \citenamefont {Schmitt-Rink}}]{PhysRevB.46.7252}%
  \BibitemOpen
  \bibfield  {author} {\bibinfo {author} {\bibfnamefont {J.}~\bibnamefont
  {Feldmann}}, \bibinfo {author} {\bibfnamefont {K.}~\bibnamefont {Leo}},
  \bibinfo {author} {\bibfnamefont {J.}~\bibnamefont {Shah}}, \bibinfo {author}
  {\bibfnamefont {D.~A.~B.}\ \bibnamefont {Miller}}, \bibinfo {author}
  {\bibfnamefont {J.~E.}\ \bibnamefont {Cunningham}}, \bibinfo {author}
  {\bibfnamefont {T.}~\bibnamefont {Meier}}, \bibinfo {author} {\bibfnamefont
  {G.}~\bibnamefont {von Plessen}}, \bibinfo {author} {\bibfnamefont
  {A.}~\bibnamefont {Schulze}}, \bibinfo {author} {\bibfnamefont
  {P.}~\bibnamefont {Thomas}},\ and\ \bibinfo {author} {\bibfnamefont
  {S.}~\bibnamefont {Schmitt-Rink}},\ }\bibinfo {title} {Optical investigation
  of Bloch oscillations in a semiconductor superlattice},\ \href
  {https://doi.org/10.1103/PhysRevB.46.7252} {\bibfield  {journal} {\bibinfo
  {journal} {Phys. Rev. B}\ }\textbf {\bibinfo {volume} {46}},\ \bibinfo
  {pages} {7252} (\bibinfo {year} {1992})}\BibitemShut {NoStop}%
\bibitem [{\citenamefont {Ben~Dahan}\ \emph {et~al.}(1996)\citenamefont
  {Ben~Dahan}, \citenamefont {Peik}, \citenamefont {Reichel}, \citenamefont
  {Castin},\ and\ \citenamefont {Salomon}}]{PhysRevLett.76.4508}%
  \BibitemOpen
  \bibfield  {author} {\bibinfo {author} {\bibfnamefont {M.}~\bibnamefont
  {Ben~Dahan}}, \bibinfo {author} {\bibfnamefont {E.}~\bibnamefont {Peik}},
  \bibinfo {author} {\bibfnamefont {J.}~\bibnamefont {Reichel}}, \bibinfo
  {author} {\bibfnamefont {Y.}~\bibnamefont {Castin}},\ and\ \bibinfo {author}
  {\bibfnamefont {C.}~\bibnamefont {Salomon}},\ }\bibinfo {title} {Bloch
  Oscillations of Atoms in an Optical Potential},\ \href
  {https://doi.org/10.1103/PhysRevLett.76.4508} {\bibfield  {journal} {\bibinfo
   {journal} {Phys. Rev. Lett.}\ }\textbf {\bibinfo {volume} {76}},\ \bibinfo
  {pages} {4508} (\bibinfo {year} {1996})}\BibitemShut {NoStop}%
\bibitem [{\citenamefont {Guo}\ \emph {et~al.}(2021)\citenamefont {Guo},
  \citenamefont {Ge}, \citenamefont {Li}, \citenamefont {Wang}, \citenamefont
  {Zhang}, \citenamefont {Song}, \citenamefont {Xiang}, \citenamefont {Song},
  \citenamefont {Jin}, \citenamefont {Lu} \emph {et~al.}}]{guo2021observation}%
  \BibitemOpen
  \bibfield  {author} {\bibinfo {author} {\bibfnamefont {X.-Y.}\ \bibnamefont
  {Guo}}, \bibinfo {author} {\bibfnamefont {Z.-Y.}\ \bibnamefont {Ge}},
  \bibinfo {author} {\bibfnamefont {H.}~\bibnamefont {Li}}, \bibinfo {author}
  {\bibfnamefont {Z.}~\bibnamefont {Wang}}, \bibinfo {author} {\bibfnamefont
  {Y.-R.}\ \bibnamefont {Zhang}}, \bibinfo {author} {\bibfnamefont
  {P.}~\bibnamefont {Song}}, \bibinfo {author} {\bibfnamefont {Z.}~\bibnamefont
  {Xiang}}, \bibinfo {author} {\bibfnamefont {X.}~\bibnamefont {Song}},
  \bibinfo {author} {\bibfnamefont {Y.}~\bibnamefont {Jin}}, \bibinfo {author}
  {\bibfnamefont {L.}~\bibnamefont {Lu}}, \emph {et~al.},\ }\bibinfo {title}
  {Observation of Bloch oscillations and Wannier-Stark localization on a
  superconducting quantum processor},\ \href
  {https://doi.org/10.1038/s41534-021-00385-3} {\bibfield  {journal} {\bibinfo
  {journal} {npj Quantum Inf.}\ }\textbf {\bibinfo {volume} {7}},\ \bibinfo
  {pages} {51} (\bibinfo {year} {2021})}\BibitemShut {NoStop}%
\bibitem [{\citenamefont {Surace}\ \emph {et~al.}(2020)\citenamefont {Surace},
  \citenamefont {Mazza}, \citenamefont {Giudici}, \citenamefont {Lerose},
  \citenamefont {Gambassi},\ and\ \citenamefont
  {Dalmonte}}]{PhysRevX.10.021041}%
  \BibitemOpen
  \bibfield  {author} {\bibinfo {author} {\bibfnamefont {F.~M.}\ \bibnamefont
  {Surace}}, \bibinfo {author} {\bibfnamefont {P.~P.}\ \bibnamefont {Mazza}},
  \bibinfo {author} {\bibfnamefont {G.}~\bibnamefont {Giudici}}, \bibinfo
  {author} {\bibfnamefont {A.}~\bibnamefont {Lerose}}, \bibinfo {author}
  {\bibfnamefont {A.}~\bibnamefont {Gambassi}},\ and\ \bibinfo {author}
  {\bibfnamefont {M.}~\bibnamefont {Dalmonte}},\ }\bibinfo {title} {Lattice
  Gauge Theories and String Dynamics in Rydberg Atom Quantum Simulators},\
  \href {https://doi.org/10.1103/PhysRevX.10.021041} {\bibfield  {journal}
  {\bibinfo  {journal} {Phys. Rev. X}\ }\textbf {\bibinfo {volume} {10}},\
  \bibinfo {pages} {021041} (\bibinfo {year} {2020})}\BibitemShut {NoStop}%
\bibitem [{\citenamefont {Yang}\ \emph
  {et~al.}(2020{\natexlab{b}})\citenamefont {Yang}, \citenamefont {Liu},
  \citenamefont {Gorshkov},\ and\ \citenamefont
  {Iadecola}}]{PhysRevLett.124.207602}%
  \BibitemOpen
  \bibfield  {author} {\bibinfo {author} {\bibfnamefont {Z.-C.}\ \bibnamefont
  {Yang}}, \bibinfo {author} {\bibfnamefont {F.}~\bibnamefont {Liu}}, \bibinfo
  {author} {\bibfnamefont {A.~V.}\ \bibnamefont {Gorshkov}},\ and\ \bibinfo
  {author} {\bibfnamefont {T.}~\bibnamefont {Iadecola}},\ }\bibinfo {title}
  {Hilbert-Space Fragmentation from Strict Confinement},\ \href
  {https://doi.org/10.1103/PhysRevLett.124.207602} {\bibfield  {journal}
  {\bibinfo  {journal} {Phys. Rev. Lett.}\ }\textbf {\bibinfo {volume} {124}},\
  \bibinfo {pages} {207602} (\bibinfo {year} {2020}{\natexlab{b}})}\BibitemShut
  {NoStop}%
\bibitem [{\citenamefont {Smith}\ \emph {et~al.}(2017)\citenamefont {Smith},
  \citenamefont {Knolle}, \citenamefont {Kovrizhin},\ and\ \citenamefont
  {Moessner}}]{PhysRevLett.118.266601}%
  \BibitemOpen
  \bibfield  {author} {\bibinfo {author} {\bibfnamefont {A.}~\bibnamefont
  {Smith}}, \bibinfo {author} {\bibfnamefont {J.}~\bibnamefont {Knolle}},
  \bibinfo {author} {\bibfnamefont {D.~L.}\ \bibnamefont {Kovrizhin}},\ and\
  \bibinfo {author} {\bibfnamefont {R.}~\bibnamefont {Moessner}},\ }\bibinfo
  {title} {Disorder-free localization},\ \href
  {https://doi.org/10.1103/PhysRevLett.118.266601} {\bibfield  {journal}
  {\bibinfo  {journal} {Phys. Rev. Lett.}\ }\textbf {\bibinfo {volume} {118}},\
  \bibinfo {pages} {266601} (\bibinfo {year} {2017})}\BibitemShut {NoStop}%
\bibitem [{\citenamefont {Ge}\ \emph {et~al.}(2024)\citenamefont {Ge},
  \citenamefont {Zhang},\ and\ \citenamefont {Nori}}]{PhysRevLett.132.230403}%
  \BibitemOpen
  \bibfield  {author} {\bibinfo {author} {\bibfnamefont {Z.-Y.}\ \bibnamefont
  {Ge}}, \bibinfo {author} {\bibfnamefont {Y.-R.}\ \bibnamefont {Zhang}},\ and\
  \bibinfo {author} {\bibfnamefont {F.}~\bibnamefont {Nori}},\ }\bibinfo
  {title} {Nonmesonic Quantum Many-Body Scars in a 1D Lattice Gauge Theory},\
  \href {https://doi.org/10.1103/PhysRevLett.132.230403} {\bibfield  {journal}
  {\bibinfo  {journal} {Phys. Rev. Lett.}\ }\textbf {\bibinfo {volume} {132}},\
  \bibinfo {pages} {230403} (\bibinfo {year} {2024})}\BibitemShut {NoStop}%
\bibitem [{\citenamefont {Arute}\ \emph {et~al.}(2019)\citenamefont {Arute},
  \citenamefont {Arya}, \citenamefont {Babbush}, \citenamefont {Bacon},
  \citenamefont {Bardin}, \citenamefont {Barends}, \citenamefont {Biswas},
  \citenamefont {Boixo}, \citenamefont {Brandao}, \citenamefont {Buell},
  \citenamefont {Burkett}, \citenamefont {Chen}, \citenamefont {Chen},
  \citenamefont {Chiaro}, \citenamefont {Collins}, \citenamefont {Courtney},
  \citenamefont {Dunsworth}, \citenamefont {Farhi}, \citenamefont {Foxen},
  \citenamefont {Fowler}, \citenamefont {Gidney}, \citenamefont {Giustina},
  \citenamefont {Graff}, \citenamefont {Guerin}, \citenamefont {Habegger},
  \citenamefont {Harrigan}, \citenamefont {Hartmann}, \citenamefont {Ho},
  \citenamefont {Hoffmann}, \citenamefont {Huang}, \citenamefont {Humble},
  \citenamefont {Isakov}, \citenamefont {Jeffrey}, \citenamefont {Jiang},
  \citenamefont {Kafri}, \citenamefont {Kechedzhi}, \citenamefont {Kelly},
  \citenamefont {Klimov}, \citenamefont {Knysh}, \citenamefont {Korotkov},
  \citenamefont {Kostritsa}, \citenamefont {Landhuis}, \citenamefont
  {Lindmark}, \citenamefont {Lucero}, \citenamefont {Lyakh}, \citenamefont
  {Mandra}, \citenamefont {McClean}, \citenamefont {McEwen}, \citenamefont
  {Megrant}, \citenamefont {Mi}, \citenamefont {Michielsen}, \citenamefont
  {Mohseni}, \citenamefont {Mutus}, \citenamefont {Naaman}, \citenamefont
  {Neeley}, \citenamefont {Neill}, \citenamefont {Niu}, \citenamefont {Ostby},
  \citenamefont {Petukhov}, \citenamefont {Platt}, \citenamefont {Quintana},
  \citenamefont {Rieffel}, \citenamefont {Roushan}, \citenamefont {Rubin},
  \citenamefont {Sank}, \citenamefont {Satzinger}, \citenamefont {Smelyanskiy},
  \citenamefont {Sung}, \citenamefont {Trevithick}, \citenamefont
  {Vainsencher}, \citenamefont {Villalonga}, \citenamefont {White},
  \citenamefont {Yao}, \citenamefont {Yeh}, \citenamefont {Zalcman},
  \citenamefont {Neven},\ and\ \citenamefont {Martinis}}]{WOS:000492991700045}%
  \BibitemOpen
  \bibfield  {author} {\bibinfo {author} {\bibfnamefont {F.}~\bibnamefont
  {Arute}}, \bibinfo {author} {\bibfnamefont {K.}~\bibnamefont {Arya}},
  \bibinfo {author} {\bibfnamefont {R.}~\bibnamefont {Babbush}}, \bibinfo
  {author} {\bibfnamefont {D.}~\bibnamefont {Bacon}}, \bibinfo {author}
  {\bibfnamefont {J.~C.}\ \bibnamefont {Bardin}}, \bibinfo {author}
  {\bibfnamefont {R.}~\bibnamefont {Barends}}, \bibinfo {author} {\bibfnamefont
  {R.}~\bibnamefont {Biswas}}, \bibinfo {author} {\bibfnamefont
  {S.}~\bibnamefont {Boixo}}, \bibinfo {author} {\bibfnamefont {F.~G. S.~L.}\
  \bibnamefont {Brandao}}, \bibinfo {author} {\bibfnamefont {D.~A.}\
  \bibnamefont {Buell}}, \bibinfo {author} {\bibfnamefont {B.}~\bibnamefont
  {Burkett}}, \bibinfo {author} {\bibfnamefont {Y.}~\bibnamefont {Chen}},
  \bibinfo {author} {\bibfnamefont {Z.}~\bibnamefont {Chen}}, \bibinfo {author}
  {\bibfnamefont {B.}~\bibnamefont {Chiaro}}, \bibinfo {author} {\bibfnamefont
  {R.}~\bibnamefont {Collins}}, \bibinfo {author} {\bibfnamefont
  {W.}~\bibnamefont {Courtney}}, \bibinfo {author} {\bibfnamefont
  {A.}~\bibnamefont {Dunsworth}}, \bibinfo {author} {\bibfnamefont
  {E.}~\bibnamefont {Farhi}}, \bibinfo {author} {\bibfnamefont
  {B.}~\bibnamefont {Foxen}}, \bibinfo {author} {\bibfnamefont
  {A.}~\bibnamefont {Fowler}}, \bibinfo {author} {\bibfnamefont
  {C.}~\bibnamefont {Gidney}}, \bibinfo {author} {\bibfnamefont
  {M.}~\bibnamefont {Giustina}}, \bibinfo {author} {\bibfnamefont
  {R.}~\bibnamefont {Graff}}, \bibinfo {author} {\bibfnamefont
  {K.}~\bibnamefont {Guerin}}, \bibinfo {author} {\bibfnamefont
  {S.}~\bibnamefont {Habegger}}, \bibinfo {author} {\bibfnamefont {M.~P.}\
  \bibnamefont {Harrigan}}, \bibinfo {author} {\bibfnamefont {M.~J.}\
  \bibnamefont {Hartmann}}, \bibinfo {author} {\bibfnamefont {A.}~\bibnamefont
  {Ho}}, \bibinfo {author} {\bibfnamefont {M.}~\bibnamefont {Hoffmann}},
  \bibinfo {author} {\bibfnamefont {T.}~\bibnamefont {Huang}}, \bibinfo
  {author} {\bibfnamefont {T.~S.}\ \bibnamefont {Humble}}, \bibinfo {author}
  {\bibfnamefont {S.~V.}\ \bibnamefont {Isakov}}, \bibinfo {author}
  {\bibfnamefont {E.}~\bibnamefont {Jeffrey}}, \bibinfo {author} {\bibfnamefont
  {Z.}~\bibnamefont {Jiang}}, \bibinfo {author} {\bibfnamefont
  {D.}~\bibnamefont {Kafri}}, \bibinfo {author} {\bibfnamefont
  {K.}~\bibnamefont {Kechedzhi}}, \bibinfo {author} {\bibfnamefont
  {J.}~\bibnamefont {Kelly}}, \bibinfo {author} {\bibfnamefont {P.~V.}\
  \bibnamefont {Klimov}}, \bibinfo {author} {\bibfnamefont {S.}~\bibnamefont
  {Knysh}}, \bibinfo {author} {\bibfnamefont {A.}~\bibnamefont {Korotkov}},
  \bibinfo {author} {\bibfnamefont {F.}~\bibnamefont {Kostritsa}}, \bibinfo
  {author} {\bibfnamefont {D.}~\bibnamefont {Landhuis}}, \bibinfo {author}
  {\bibfnamefont {M.}~\bibnamefont {Lindmark}}, \bibinfo {author}
  {\bibfnamefont {E.}~\bibnamefont {Lucero}}, \bibinfo {author} {\bibfnamefont
  {D.}~\bibnamefont {Lyakh}}, \bibinfo {author} {\bibfnamefont
  {S.}~\bibnamefont {Mandra}}, \bibinfo {author} {\bibfnamefont {J.~R.}\
  \bibnamefont {McClean}}, \bibinfo {author} {\bibfnamefont {M.}~\bibnamefont
  {McEwen}}, \bibinfo {author} {\bibfnamefont {A.}~\bibnamefont {Megrant}},
  \bibinfo {author} {\bibfnamefont {X.}~\bibnamefont {Mi}}, \bibinfo {author}
  {\bibfnamefont {K.}~\bibnamefont {Michielsen}}, \bibinfo {author}
  {\bibfnamefont {M.}~\bibnamefont {Mohseni}}, \bibinfo {author} {\bibfnamefont
  {J.}~\bibnamefont {Mutus}}, \bibinfo {author} {\bibfnamefont
  {O.}~\bibnamefont {Naaman}}, \bibinfo {author} {\bibfnamefont
  {M.}~\bibnamefont {Neeley}}, \bibinfo {author} {\bibfnamefont
  {C.}~\bibnamefont {Neill}}, \bibinfo {author} {\bibfnamefont {M.~Y.}\
  \bibnamefont {Niu}}, \bibinfo {author} {\bibfnamefont {E.}~\bibnamefont
  {Ostby}}, \bibinfo {author} {\bibfnamefont {A.}~\bibnamefont {Petukhov}},
  \bibinfo {author} {\bibfnamefont {J.~C.}\ \bibnamefont {Platt}}, \bibinfo
  {author} {\bibfnamefont {C.}~\bibnamefont {Quintana}}, \bibinfo {author}
  {\bibfnamefont {E.~G.}\ \bibnamefont {Rieffel}}, \bibinfo {author}
  {\bibfnamefont {P.}~\bibnamefont {Roushan}}, \bibinfo {author} {\bibfnamefont
  {N.~C.}\ \bibnamefont {Rubin}}, \bibinfo {author} {\bibfnamefont
  {D.}~\bibnamefont {Sank}}, \bibinfo {author} {\bibfnamefont {K.~J.}\
  \bibnamefont {Satzinger}}, \bibinfo {author} {\bibfnamefont {V.}~\bibnamefont
  {Smelyanskiy}}, \bibinfo {author} {\bibfnamefont {K.~J.}\ \bibnamefont
  {Sung}}, \bibinfo {author} {\bibfnamefont {M.~D.}\ \bibnamefont
  {Trevithick}}, \bibinfo {author} {\bibfnamefont {A.}~\bibnamefont
  {Vainsencher}}, \bibinfo {author} {\bibfnamefont {B.}~\bibnamefont
  {Villalonga}}, \bibinfo {author} {\bibfnamefont {T.}~\bibnamefont {White}},
  \bibinfo {author} {\bibfnamefont {Z.~J.}\ \bibnamefont {Yao}}, \bibinfo
  {author} {\bibfnamefont {P.}~\bibnamefont {Yeh}}, \bibinfo {author}
  {\bibfnamefont {A.}~\bibnamefont {Zalcman}}, \bibinfo {author} {\bibfnamefont
  {H.}~\bibnamefont {Neven}},\ and\ \bibinfo {author} {\bibfnamefont {J.~M.}\
  \bibnamefont {Martinis}},\ }\bibinfo {title} {Quantum supremacy using a
  programmable superconducting processor},\ \href
  {https://doi.org/10.1038/s41586-019-1666-5} {\bibfield  {journal} {\bibinfo
  {journal} {Nature}\ }\textbf {\bibinfo {volume} {574}},\ \bibinfo {pages}
  {505} (\bibinfo {year} {2019})}\BibitemShut {NoStop}%
\end{thebibliography}%

\clearpage 
\onecolumngrid
\begin{center}
{\large \bf Supplemental Material for:\\Observation of Inelastic Meson Scattering in a Floquet System using a Digital Quantum Simulator}
\\
\vspace{0.3cm}
\end{center}

\vspace{0.6cm}

\beginsupplement
\makeatletter
\def\@hangfrom@section#1#2#3{\@hangfrom{#1#2#3}}
\makeatother

\setcounter{section}{0}
\renewcommand{\thesection}{Supplementary Note \arabic{section}}
\renewcommand{\figurename}{Supplementary Fig.}
\setcounter{figure}{0}  

\renewcommand{\tablename}{Supplementary Table}
\setcounter{table}{0}
\renewcommand {\thetable} {\arabic{table}}%
\tableofcontents

\begin{figure*}[b!]
	\centering
	\includegraphics[width=0.99\linewidth]{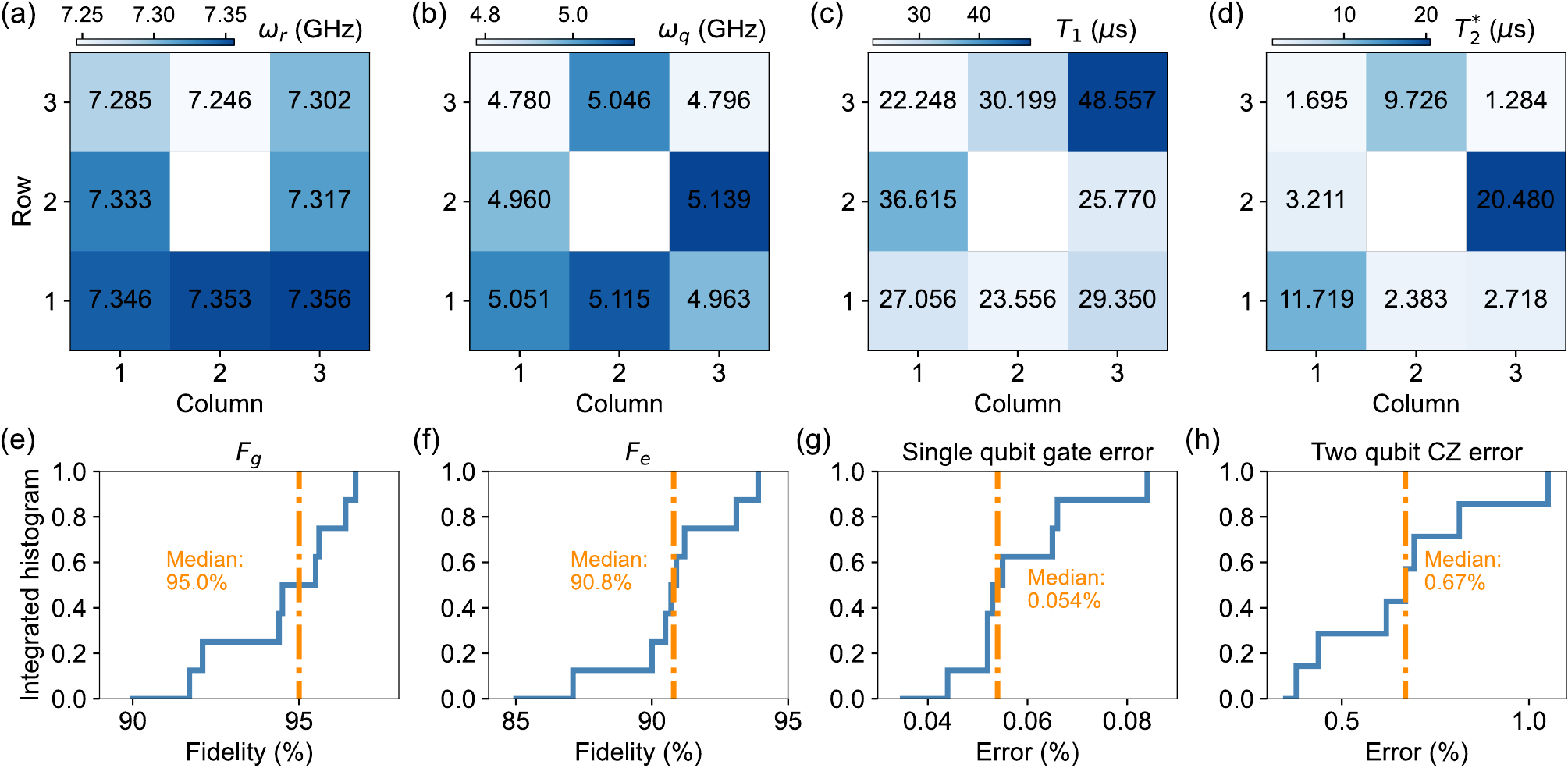}
	\caption{{Device parameters and gate fidelities.} (a)  The readout resonator frequency $\omega_{r}$. (b) Qubit idle frequency $\omega_{q}$. (c, d) Qubit relaxation time $T_1$ and dephasing time $T_2^{\ast}$ measured at the idle frequency. (e) Readout fidelity of the qubit $|0\rangle$ state. (f) Readout fidelity of the qubit $|1\rangle$ state. (g) Pauli errors of the single-qubit gates. (h) Pauli errors of the two-qubit CZ gates. The Pauli errors for both the single-qubit gates and the two-qubit CZ gates are measured via randomized benchmarking.}
	\label{fig:parameters}
\end{figure*}
\section{Device information}
The superconducting processor employed in this work consists of 9 transmon qubits (3×3 square lattice) and 12 couplers, fabricated using flip-chip technology~\cite{WOS:000492991700045}  with a tantalum base layer~\cite{fab,fab2} (schematic shown in Fig. 1 in the main text). To implement the experimental protocol, we decouple Q$_0$ from Q$_7$ and isolate the central qubit from its neighboring qubits.

Our experiments employed eight qubits (Q$_0$ to Q$7$; see Fig.~1 in the main text), with key device parameters presented in Supplementary Fig.~\ref{fig:parameters}. The readout resonator frequencies $\omega_r$, ranging from 7.246 GHz to 7.353 GHz, exceed the maximum qubit frequencies. Qubits' idle frequencies $\omega{_q}$ are arranged close to the maximum qubit frequencies to mitigate flux noise effects. At the idle point, the energy relaxation times $T_1$ exceed 22 $\mu$s (this timescale is much longer than the characteristic runtime of our quantum circuit) for all qubits, while dephasing times $T_2^{\ast}$ range from 1.60 $\mu$s to 20.5 $\mu$s.
Through comprehensive signal calibration and error mitigation protocols, we demonstrate high-fidelity gate operations with single-qubit Pauli errors below 0.1$\%$ and a median two-qubit CZ gate error of $\sim$0.67$\%$. For detailed information about the measurement setup, device parameters, and calibration procedures, please refer to Refs.~\cite{wang_quantum_2024,wang_demonstration_2024}.

For quantum simulation of key phenomena in lattice gauge theories, including confinement dynamics and string breaking, we prepared a variational set of initial states, including $|10000000\rangle$, $|00010000\rangle$, $|00100100\rangle$, $|00011000\rangle$, $|11110000\rangle$ and $|00111100\rangle$. The fidelity of these initial states are measured using quantum state tomography, with the results displayed in Supplementary Fig.~\ref{fig:initialstatefidelity}.

\begin{figure*}[htbp!]
    \centering
    \includegraphics[width=0.8\linewidth]{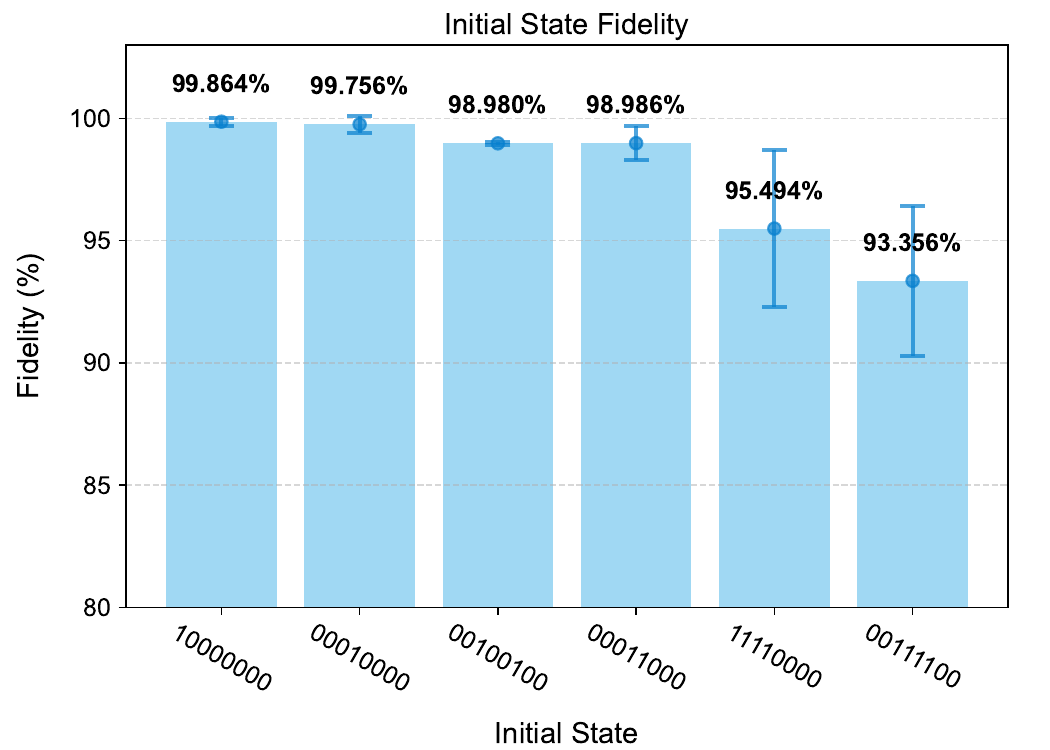}
    \caption{Fidelities of various initial states measured via quantum state tomography. Error bars represent the one standard deviation.}
    \label{fig:initialstatefidelity}
\end{figure*}

\section{Effective lattice gauge theory}

In the experiment, we consider a Floquet dynamics governed by a unitary operator
 \begin{align}
  \hat{U} = \exp(-i h \hat{Z})\exp(-i \mu \hat{X}) \exp(i J \hat{H}_{zz}),
\end{align}
where 
 \begin{align} 
\hat{Z} = \sum_j \hat{\sigma}_j^z, \ \ \ \hat{X} = \sum_j \hat{\sigma}_j^x, \ \ \  \hat{H}_{zz} = -\sum_j \hat{\sigma}_j^z \hat{\sigma}_{j+1}^z.
\end{align}
To map it to a LGT coupled to a matter field, we first use the bond index to replace the site index: $j\rightarrow j+1/2$, where we have
 \begin{align} \label{U1}
	\hat{Z} = \sum_j \hat{\sigma}_{j+1/2}^z, \ \ \ \hat{X} = \sum_j \hat{\sigma}_{j+1/2}^x, \ \ \  \hat{H}_{zz} = -\sum_j \hat{\sigma}_{j-1/2}^z \hat{\sigma}_{j+1/2}^z.
\end{align}
introduce the following transformation:
\begin{equation}
\hat \sigma_{j+1/2}^z = \hat \tau^x_{j+1/2} , \ \ \ \ \ \hat\sigma^x_{j+1/2} =  \hat{s}^x_j \hat \tau_{j+1/2}^z \hat s^x_{j+1}.
\end{equation}
When $\hat{s}_j^\alpha$ ($\alpha = x, y, z$) is a Pauli matrix, then we can verify  that  $\hat{\tau}_j^\alpha$  also satisfy the commutation relation of Pauli matrices.
We label  $\hat{s}_j^\alpha$ and $\hat{\tau}_j^\alpha$ ($\alpha = x, y, z$)  as the matter and gauge fields, respectively.
  
Thus,  the terms in Eq.~(\ref{U1}) can be written as 
 \begin{align} 
	\hat{Z} = \sum_j \hat{\tau}_{j+1/2}^x, \ \ \ \hat{X} = \sum_j \hat{s}^x_j \hat \tau_{j+1/2}^z \hat s^x_{j+1}, \ \ \  \hat{H}_{zz} = -\sum_j \hat{\tau}_{j-1/2}^x\hat{\tau}_{j+1/2}^x.
\end{align}
We can find that the operators $\hat{Z}$, $\hat{Z}$, and $\hat{H}_{zz}$ all commute to a $\mathbb{Z}_2$  gauge generator
 \begin{align} 
\hat G_{j} := \hat \tau^x_{j-\frac{1}{2}} \hat s_j^z\hat \tau^x_{j+\frac{1}{2}} .
\end{align}
Here, when fixing the gauge sector $\hat G_j = 1$, we have
 \begin{align} 
 \hat{H}_{zz} = -\sum_j \hat s^z_{j}.
\end{align}
Thus, the Floquet dynamics of this spin chain can be mapped to a Floquet $\mathbb{Z}_2$ LGT
 \begin{align} \label{Ulgt_2}
	\hat{U} = &\exp({-i h\sum_j \hat{\tau}_{j+1/2}^x})\exp( {-i\mu\sum_{j}\hat s_j^x \hat \tau^z_{j+\frac{1}{2}} \hat s_{j+1}^x})
\exp( iJ\sum_{j}\hat s_j^z),
\end{align}
where the first term is an electric field, the second term is a kinetic term, and the third term is a mass term.

\end{document}